\documentclass[onecolumn]{article}%
\usepackage{amssymb}
\usepackage{amsmath}
\usepackage{amsfonts}
\usepackage{graphicx}%
\begin{document}

\title{\hfill{\normalsize UMTG - 1}\bigskip\\Biorthogonal Quantum Systems}
\author{Thomas Curtright$^{\text{\S ,\dag,}}${\small *} and Luca
Mezincescu$^{\text{\S ,}}${\small *}{\large \medskip}\\$^{\text{\S }}${\small Department of Physics, University of Miami, Coral
Gables, Florida 33124}\\$^{\text{\dag}}${\small School of Natural Sciences, Institute for Advanced
Study, Princeton, New Jersey 08540}}
\maketitle

\begin{abstract}
Models of PT symmetric quantum mechanics provide examples of biorthogonal
quantum systems. \ The latter incorporporate all the structure of PT symmetric
models, and allow for generalizations, especially in situations where the PT
construction of the dual space fails. \ The formalism is illustrated by a few
exact results for models of the form $H=\left(  p+\nu\right)  ^{2}+\sum
_{k>0}\mu_{k}\exp\left(  ikx\right)  $. \ In some non-trivial cases,
equivalent hermitian theories are obtained and shown to be very simple: \ They
are just free (chiral) particles. \ Field theory extensions are briefly considered.

\vfill

\noindent\underline
{$\ \ \ \ \ \ \ \ \ \ \ \ \ \ \ \ \ \ \ \ \ \ \ \ \ \ \ \ \ \ \ \ \ \ \ \ \ \ \ \ \ \ \ \ \ \ \ \ \ \ \ \ \ \ \ $%
}

{\small *}curtright@physics.miami.edu \& mezincescu@physics.miami.edu

\end{abstract}

\section{Introduction}

There has been some recent theoretical interest in non-hermitian
Schr\"{o}dinger equations, especially in the guise of \textquotedblleft PT
symmetric theories\textquotedblright\ \cite{BenderReview,MostaReview}.
\ Therefore, we have considered a few simple soluble examples and explored
them in some detail. \ We believe exact solvability permits the underlying
structure to be appreciated more completely. To this end we have found
Neumann's work \cite{Neumann} to be remarkably prescient. \ Although in
physics there has long been an interest in complex Hamiltonians for
phenomenological purposes \cite{Feshbach,Wong,Faisal}, quite generally
mathematical interest in non-self-adjoint spectral problems pre-dates the
invention of quantum mechanics \cite{Birkhoff}. \ See \cite{Dunford}\ for a
brief survey of the early mathematical history.

Some previous papers have touched on the relevance of biorthogonal systems for
PT symmetric models \cite{MostaJMP2,MostaJMP3,MostaJMP4,MostaBatal,Znojil}.
\ Here we wish to stress the importance of such systems, and their generality,
in the context of elementary soluble cases. \ We consider models whose
Hamiltonians are of the form
\cite{McGarvey,Rofe,Gasymov,Veliev,Birnir,Pastur,Deift,Shin}%
\begin{equation}
H=\left(  p+\nu\right)  ^{2}+\sum_{k>0}\mu_{k}\exp\left(  ikx\right)
\label{H}%
\end{equation}
where the imaginary exponents are of the \emph{same} sign, and where $\nu$ and
$\mu_{k}$ have arbitrary values. \ Although much is known about the
mathematical structure of these models, they tend to be overlooked in the
physics literature on PT symmetric theories. \ We seek to remedy this by
offering a thorough discussion of their properties, especially in the case of
periodic solutions, for which we believe we make several new observations.
\ We analyze at considerable length the cases of one and two exponentials, and
then we indicate how the results extend to the general case. \ We believe it
is almost impossible to over-emphasize the merits of going through such
exactly solvable models in complete detail -- \textquotedblleft the simpler
the better\textquotedblright\ if the important structural features are
retained. \ We end with a brief look at field theoretic extensions.

\section{Biorthogonal systems}

A sequence of elements $\left\{  \psi_{j}\right\}  $ and linear functionals
$\left\{  \Lambda_{k}\right\}  $ is said to be \emph{biorthogonal} (or more
precisely, \emph{biorthonormal}) when%
\begin{equation}
\Lambda_{k}\left(  \psi_{j}\right)  =\delta_{jk}%
\end{equation}
See the classic texts by Goursat (\cite{Goursat} Vol. III \S 573 and \S 594),
Banach (\cite{Banach} Chapter 7), Morse and Feshbach (\cite{Morse} \S 7.5 and
\S 8.3), and Gohberg and Krein (\cite{Gogh} Chapter 6), as well as
\cite{Dieudonne,Bari,Muskhelishvili,Brezinski,Veliev}. \ The notion of
biorthogonality is especially useful when dealing with elements that are
analytic functions defined on a contour in the complex plane. \ In this case,
the linear functionals are usually realized as integrals along the contour
involving a set of \textquotedblleft dual functions\textquotedblright%
\ $\chi_{k}\equiv\left(  \psi_{k}\right)  _{\text{dual}}$, so%
\begin{equation}
\Lambda_{k}\left(  \psi_{j}\right)  =\int\chi_{k}\left(  z\right)  \psi
_{j}\left(  z\right)  dz=\delta_{jk} \label{Biorthogonality}%
\end{equation}
However, it is important to stress that \emph{a priori} there need not be a
simple point-by-point relation between the functions and their duals (unlike
the standard situation in quantum mechanics). \ In general, $\chi_{k}\left(
z\right)  $ is \emph{not} obtained just by conjugation, and differs
substantially from $\overline{\psi_{j}\left(  z\right)  }$, even when the
contour of integration is a real line segment. \ Moreover, the normalization
and phase of $\psi_{j}$ can be changed, arbitrarily, if a compensating change
is made in the normalization and phase of $\chi_{j}$. \ That is to say, if the
system $\left\{  \psi_{j},\chi_{k}\right\}  $ satisfies (\ref{Biorthogonality}%
), then so does the system $\left\{  Z_{j}\psi_{j},\chi_{k}/Z_{k}\right\}  $
(no sum on $j,k$) for all non-vanishing, finite choices of the $Z$s.

For any function in the span of $\left\{  \psi_{j}\right\}  $, we have%
\begin{equation}
\psi\left(  z\right)  =\sum_{n}c_{n}\psi_{n}\left(  z\right)  \ ,\ \ \ c_{n}%
=\int\chi_{n}\left(  z\right)  \psi\left(  z\right)  dz \label{Psi}%
\end{equation}
assuming the requisite convergence of the sum. \ For such a $\psi\left(
z\right)  $ we may define, at least formally, an associated dual function by
\begin{equation}
\psi_{\text{dual}}\left(  z\right)  \equiv\sum_{k}c_{k}^{\ast}\chi_{k}\left(
z\right)  \label{PsiDual}%
\end{equation}
While the convergence of the sum for $\psi_{\text{dual}}\left(  z\right)
$\ is not necessarily guaranteed by the convergence of that for $\psi\left(
z\right)  $, nevertheless, if it is permissible to interchange $\sum_{k}$ and
$\int dz$, then we can make use of (\ref{Psi}) and define a norm as%
\begin{equation}
\left\Vert \psi\right\Vert ^{2}\equiv\int\psi_{\text{dual}}\left(  z\right)
\psi\left(  z\right)  \ dz=\sum_{n}\left\vert c_{n}\right\vert ^{2}>0
\label{Norm}%
\end{equation}
If $\left\Vert \psi\right\Vert ^{2}$ is finite, this allows us to define a
bona fide probability distribution on $n$ corresponding to the function $\psi
$. \ The probability of $n$ is just $\left\vert c_{n}\right\vert
^{2}/\left\Vert \psi\right\Vert ^{2}$. \ However, it is important to stress
that the bilinear $\psi_{\text{dual}}\left(  z\right)  \psi\left(  z\right)  $
might \emph{not} provide a bona fide probability distribution in the variable
$z$, since $\psi_{\text{dual}}\left(  z\right)  \psi\left(  z\right)  $ might
be complex. \ Moreover, $\int\psi_{\text{dual}}\left(  z\right)
\ z\ \psi\left(  z\right)  dz$ might be complex for one or both of two
reasons: \ $z $ might be on a complex contour, and $\psi_{\text{dual}}\left(
z\right)  $ is not necessarily $\overline{\psi\left(  z\right)  }$, in
general. \ (Perhaps it is possible to think of $\psi_{\text{dual}}\left(
z\right)  \psi\left(  z\right)  $\ as a \textquotedblleft complex
probability\textquotedblright\ density \cite{Dirac,Youssef} but we will not
pursue that idea here.)

When $n$ labels the eigenvalue of an operator $L$, such that $L\psi_{n}\left(
z\right)  =\lambda_{n}\psi_{n}\left(  z\right)  $, we may also compute
averages\ of functions $f\left(  L\right)  $ using a procedure analogous to
that of conventional quantum mechanics, even if $\psi_{\text{dual}}\left(
z\right)  \neq\overline{\psi\left(  z\right)  }$. \ These statistical averages
are defined to be
\begin{equation}
\left\langle f\right\rangle _{\psi}=\frac{1}{\left\Vert \psi\right\Vert ^{2}%
}\int\psi_{\text{dual}}\left(  z\right)  \ f\left(  L\right)  \ \psi\left(
z\right)  \ dz=\sum_{n}\left\vert c_{n}\right\vert ^{2}f\left(  \lambda
_{n}\right)  /\left\Vert \psi\right\Vert ^{2} \label{fAverage}%
\end{equation}
again assuming the requisite convergences. \ Whether these averages are real
depends on both $f$ and $\left\{  \lambda_{n}\right\}  $. \ Also, for any
given $L$, the action of the operator on $\psi_{\text{dual}}\left(  z\right)
$ is not necessarily obvious from its action on $\psi$. \ Similar statements
apply if we interchange the r\^{o}les of $\psi$ and $\psi_{\text{dual}}$.

For any biorthogonal system, it is always possible, at least formally, to
express the norms (\ref{Norm}) for functions in the span of $\left\{  \psi
_{j}\right\}  $ as integrals of bilinears in $\psi$ and $\psi^{\ast}$, where
the construction is \emph{bilocal}, in the general situation. \ If we build a
kernel from the dual basis functions as follows%
\begin{equation}
K\left(  \overline{w},z\right)  \equiv\sum_{n}\overline{\chi_{n}\left(
w\right)  }\ \chi_{n}\left(  z\right)  \label{Kernel}%
\end{equation}
once again assuming the requisite convergences, we then have%
\begin{align}
\psi_{\text{dual}}\left(  z\right)   &  =\int\overline{\psi\left(  w\right)
}\ K\left(  \overline{w},z\right)  \ d\overline{w}\\
\left\Vert \psi\right\Vert ^{2}  &  =%
%TCIMACRO{\diint }%
%BeginExpansion
{\displaystyle\iint}
%EndExpansion
\overline{\psi\left(  w\right)  }\ K\left(  \overline{w},z\right)
\ \psi\left(  z\right)  \ d\overline{w}dz \label{BiNorm}%
\end{align}
where the contour for $\overline{w}$ is the complex conjugate of that for $z$.
\ Similar results follow, at least formally, from constructing a dual bilocal
kernel and integrating this with the associated dual functions.
\begin{align}
J\left(  \overline{w},z\right)   &  \equiv\sum_{n}\overline{\psi_{n}\left(
w\right)  }\ \psi_{n}\left(  z\right) \label{DualKernel}\\
\overline{\psi\left(  w\right)  }  &  =\int J\left(  \overline{w},z\right)
\ \psi_{\text{dual}}\left(  z\right)  \ dz\\
\left\Vert \psi\right\Vert ^{2}  &  =%
%TCIMACRO{\diint }%
%BeginExpansion
{\displaystyle\iint}
%EndExpansion
\overline{\psi_{\text{dual}}\left(  w\right)  }\ J\left(  \overline
{w},z\right)  \ \psi_{\text{dual}}\left(  z\right)  \ d\overline{w}dz
\label{DualBiNorm}%
\end{align}
Also, note that the convergences of the sums for $J$ and $K$ can be affected
by making compensating changes in the individual normalizations of the
elements in the biorthogonal system: $\left\{  \psi_{j},\chi_{k}\right\}
\rightarrow\left\{  Z_{j}\psi_{j},\chi_{k}/Z_{k}\right\}  $. \ By suitable
choice of the $Z$s, the sum for one of the kernels can be made to converge
uniformly, and rapidly. \ In the process, the sum for the other kernel may
well become a divergent (perhaps asymptotic) series. \ 

Formal expressions for other kernels follow immediately from acting with
various operators on the $\left\{  \psi_{j}\right\}  $ or $\left\{  \chi
_{k}\right\}  $ functions, or their conjugates, under the sums. \ For
instance, returning to the previous eigenvalue example,
\begin{equation}
J_{f\left(  L\right)  }\left(  \overline{w},z\right)  \equiv\sum_{n}%
\overline{\psi_{n}\left(  w\right)  }\ f\left(  L\right)  \ \psi_{n}\left(
z\right)  =\sum_{n}\overline{\psi_{n}\left(  w\right)  }\ f\left(  \lambda
_{n}\right)  \ \psi_{n}\left(  z\right)  \label{JfKernel}%
\end{equation}
which allows writing the average (\ref{fAverage})\ as%
\begin{equation}
\left\langle f\right\rangle _{\psi}=\frac{1}{\left\Vert \psi\right\Vert ^{2}}%
%TCIMACRO{\diint }%
%BeginExpansion
{\displaystyle\iint}
%EndExpansion
\overline{\psi_{\text{dual}}\left(  w\right)  }\ J_{f\left(  L\right)
}\left(  \overline{w},z\right)  \ \psi_{\text{dual}}\left(  z\right)
\ d\overline{w}dz \label{fAverageBiLocal}%
\end{equation}
In the $J_{f\left(  L\right)  }\left(  \overline{w},z\right)  $\ form,
hermitian kernels leading to real averages are easily recognized when
$f\left(  \lambda_{n}\right)  =\overline{f\left(  \lambda_{n}\right)  }$ for
all $n$. \ Similar statements hold for kernels $K_{g\left(  L_{\text{dual}%
}\right)  }$ built of bilinears in the dual functions, when the $\chi
_{n}\left(  z\right)  $ are eigenfunctions of operators $L_{\text{dual}}$.

\paragraph{Examples}

For an elementary example, let $\left\{  \psi_{k}\right\}  =\left\{
z^{k}\ |\ k=0,1,2,\cdots\right\}  $, i.e. the usual basis for analytic
functions near the origin, and take the contour to be any simple closed curve
surrounding $z=0$. \ Then $\chi_{k}\left(  z\right)  =z^{-k-1}/2\pi i$, for
$k=0,1,2,\cdots$. \ If the contour is a circle of fixed radius $R$, with
$z=R\exp\left(  -i\theta\right)  $, and if for convenience we incorporate
various factors into a measure, $d\mu=-dz/2\pi iz=d\theta/2\pi$, then the
previous sequence $\left\{  z^{k}\right\}  $ becomes the familiar set of
left-moving eigenfunctions of $p_{\theta}=-i\partial/\partial\theta$,
$\left\{  \psi_{k}\left(  \theta\right)  \right\}  =\left\{  R^{k}\exp\left(
-ik\theta\right)  \ |\ k=0,1,\cdots\right\}  $, and the dual functions become
(upon removal of $1/z$) the right-movers, $\left\{  \chi_{k}\left(
\theta\right)  \right\}  =\left\{  \frac{\exp\left(  ik\theta\right)  }{R^{k}%
}\ |\ k=0,1,\cdots\right\}  $, so that
\begin{equation}
\frac{1}{2\pi}\int_{0}^{2\pi}\chi_{k}\left(  \theta\right)  \psi_{j}\left(
\theta\right)  d\theta=\delta_{kj}%
\end{equation}
But even in this familiar example, as defined, $\chi_{k}\left(  \theta\right)
\neq\psi_{k}^{\ast}\left(  \theta\right)  $ unless $R=1$. \ The bilocal norm
kernels are easily constructed for this elementary case: \ $J\left(
\overline{\theta},\theta\right)  =\sum_{n=0}^{\infty}R^{2n}\exp in\left(
\overline{\theta}-\theta\right)  =1/\left(  1-R^{2}\exp i\left(
\overline{\theta}-\theta\right)  \right)  $; $K\left(  \overline{\theta
},\theta\right)  =\sum_{n=0}^{\infty}R^{-2n}\exp in\left(  \theta
-\overline{\theta}\right)  =R^{2}/\left(  R^{2}-\exp i\left(  \theta
-\overline{\theta}\right)  \right)  $. \ As functions of $R$, the
complementary convergence properties of $J$\ and $K$\ are manifest.\ \ 

A more interesting example was discovered by Neumann \cite{Neumann} in his
work on Bessel functions (see \S 9.1-17 in \cite{Watson}). \ This example is
relevant to our discussion of PT invariant Hamiltonians to follow. \ For all
analytic Bessel functions of non-negative integer index%
\begin{equation}
J_{n}\left(  z\right)  =\left(  \frac{z}{2}\right)  ^{n}\sum_{k=0}^{\infty
}\frac{\left(  -1\right)  ^{k}}{k!\left(  k+n\right)  !}\left(  \frac{z}%
{2}\right)  ^{2k} \label{BesselSeries}%
\end{equation}
there are corresponding associated \emph{Neumann polynomials} $\left\{
A_{n}\right\}  $ in powers of $1/z$ that are dual to $\left\{  J_{n}\right\}
$ on any contour enclosing the origin. \ These are given by%
\begin{equation}
A_{0}\left(  z\right)  =1\ ,\ \ \ A_{1}\left(  z\right)  =\frac{2}%
{z}\ ,\ \ \ A_{n\geq2}\left(  z\right)  =n\left(  \frac{2}{z}\right)  ^{n}%
\sum_{k=0}^{\left\lfloor n/2\right\rfloor }\frac{\left(  n-k-1\right)  !}%
{k!}\left(  \frac{z}{2}\right)  ^{2k} \label{NeumannPolySeries}%
\end{equation}
For convenience we have modified the usual form of the associated polynomials,
$O_{n}$ (as given in \cite{Neumann,Watson,Abram}), and defined $A_{n}\left(
z\right)  =\varepsilon_{n}\ z\ O_{n}\left(  z\right)  $ where $\varepsilon
_{0}=1$ and $\varepsilon_{n}=2$ for $n\neq0$. \ Upon integrating
counterclockwise once around $z=0$,%
\begin{equation}
\frac{1}{2\pi i}%
%TCIMACRO{\doint }%
%BeginExpansion
{\displaystyle\oint}
%EndExpansion
\frac{dz}{z}\ A_{j}\left(  z\right)  J_{k}\left(  z\right)  =\delta_{jk}%
\end{equation}
so that $\left\{  J_{n}\left(  z\right)  \right\}  $ and $\left\{
A_{n}\left(  z\right)  \right\}  $ constitute a biorthogonal system on the
complex plane in the same sense as $\left\{  z^{n}\right\}  $ and $\left\{
z^{-n}\right\}  $, but with the very significant difference that $A_{n}\left(
z\right)  \neq\overline{J_{n}\left(  z\right)  }$ even when $z$ is on the unit
circle. \ 

There are invertible maps between these two biorthogonal systems, $\left\{
z^{j},z^{-k}\right\}  \longleftrightarrow\left\{  J_{j}\left(  z\right)
,A_{k}\left(  z\right)  \right\}  $, about which we will say more later. \ The
existence of such maps can be inferred from (\ref{BesselSeries}) and
(\ref{NeumannPolySeries}) as well as the reversions of those series to obtain
all powers of $z$.%
\begin{equation}
1=J_{0}\left(  z\right)  +2\sum_{n=1}^{\infty}J_{2n}\left(  z\right)
\ ,\ \ \ z^{k}=2^{k}\sum_{n=0}^{\infty}\frac{\left(  k+2n\right)  \left(
k+n-1\right)  !}{n!}J_{k+2n}\left(  z\right)  \text{ \ \ where \ \ }%
k=1,2,3,\cdots\label{Schlomilch}%
\end{equation}%
\begin{equation}
\frac{1}{z^{2j}}=\frac{1}{2^{2j}}\sum_{k=0}^{j}\frac{\left(  -\right)
^{j-k}A_{2k}\left(  z\right)  }{\left(  j-k\right)  !\left(  j+k\right)
!}\ ,\ \ \ \frac{1}{z^{2j-1}}=\frac{1}{2^{2j-1}}\sum_{k=1}^{j}\frac{\left(
-\right)  ^{j-k}A_{2k-1}\left(  z\right)  }{\left(  j-k\right)  !\left(
j-1+k\right)  !} \label{PowersExpandedInAs}%
\end{equation}

\paragraph{Some bilocal kernels in closed form \ }

Neumann found that the Cauchy kernel is given by (\cite{Watson}, \S 9.1)%
\begin{equation}
\frac{1}{w-z}=\frac{1}{w}\sum_{n=0}^{\infty}A_{n}\left(  w\right)
J_{n}\left(  z\right)  \label{Cauchy}%
\end{equation}
When rewritten as%
\begin{equation}
\mathsf{Id}\left(  w,z\right)  \equiv\frac{z}{z-w}=\sum_{n=0}^{\infty}%
J_{n}\left(  w\right)  A_{n}\left(  z\right)  \label{CauchyIdent}%
\end{equation}
this kernel acts as the identity when integrated on the space of Bessel
functions (or alternatively on Neumann polynomials) provided the contours of
integration are appropriately chosen to handle the singularity at $w=z$. \
\begin{equation}
J_{k}\left(  w\right)  =\frac{1}{2\pi i}%
%TCIMACRO{\doint _{\left\vert z\right\vert >\left\vert w\right\vert }}%
%BeginExpansion
{\displaystyle\oint_{\left\vert z\right\vert >\left\vert w\right\vert }}
%EndExpansion
\frac{dz}{z}\ \mathsf{Id}\left(  w,z\right)  J_{k}\left(  z\right)
\ ,\ \ \ A_{k}\left(  z\right)  =\frac{1}{2\pi i}%
%TCIMACRO{\doint _{\left\vert w\right\vert <\left\vert z\right\vert }}%
%BeginExpansion
{\displaystyle\oint_{\left\vert w\right\vert <\left\vert z\right\vert }}
%EndExpansion
\frac{dw}{w}\ A_{k}\left(  w\right)  \mathsf{Id}\left(  w,z\right)
\end{equation}
Acting with Bessel's operator on this kernel leads in turn to an example of a
bilocal eigenoperator kernel.%
\begin{align}
\mathsf{H}\left(  w,z\right)   &  \equiv\left(  w^{2}\frac{d^{2}}{dw^{2}%
}+w\frac{d}{dw}+w^{2}\right)  \mathsf{Id}\left(  w,z\right)  =\frac{2w^{2}%
z}{\left(  z-w\right)  ^{3}}+\frac{wz}{\left(  z-w\right)  ^{2}}+\frac{w^{2}%
z}{z-w}\nonumber\\
&  =\sum_{n=0}^{\infty}n^{2}J_{n}\left(  w\right)  A_{n}\left(  z\right)
\label{CauchyH}%
\end{align}%
\begin{equation}
k^{2}J_{k}\left(  w\right)  =\frac{1}{2\pi i}%
%TCIMACRO{\doint _{\left\vert z\right\vert >\left\vert w\right\vert }}%
%BeginExpansion
{\displaystyle\oint_{\left\vert z\right\vert >\left\vert w\right\vert }}
%EndExpansion
\frac{dz}{z}\ \mathsf{H}\left(  w,z\right)  J_{k}\left(  z\right)
\ ,\ \ \ k^{2}A_{k}\left(  z\right)  =\frac{1}{2\pi i}%
%TCIMACRO{\doint _{\left\vert w\right\vert <\left\vert z\right\vert }}%
%BeginExpansion
{\displaystyle\oint_{\left\vert w\right\vert <\left\vert z\right\vert }}
%EndExpansion
\frac{dw}{w}\ A_{k}\left(  w\right)  \mathsf{H}\left(  w,z\right)
\end{equation}

Neumann also showed that (\cite{Watson}, \S 11.1-2)%
\begin{equation}
J_{0}\left(  \sqrt{w^{2}+z^{2}-2wz\cos\phi}\right)  =\sum_{n=-\infty}^{\infty
}J_{n}\left(  w\right)  J_{n}\left(  z\right)  \cos\left(  n\phi\right)
\label{AdditionTheorem}%
\end{equation}
where $J_{-n}\left(  z\right)  =\left(  -1\right)  ^{n}J_{n}\left(  z\right)
$ and $J_{0}\left(  s\right)  =\sum_{k=0}^{\infty}\frac{\left(  -1\right)
^{k}}{\left(  k!\right)  ^{2}}\left(  \frac{s^{2}}{4}\right)  ^{k}$. \ From
(\ref{AdditionTheorem})\ we obtain an explicit and closed form for the dual
kernel used in the construction of the bilocal norm for this biorthogonal
system, as in (\ref{DualKernel}).%
\begin{equation}
J\left(  \overline{w},z\right)  =\sum_{n=0}^{\infty}J_{n}\left(  \overline
{w}\right)  J_{n}\left(  z\right)  =\frac{1}{2}\left(  J_{0}\left(
\overline{w}-z\right)  +J_{0}\left(  \overline{w}\right)  J_{0}\left(
z\right)  \right)  \label{AddThm}%
\end{equation}
With the normalizations chosen, the sum for the other kernel $K\left(
\overline{w},z\right)  =\sum_{n=0}^{\infty}A_{n}\left(  \overline{w}\right)
A_{n}\left(  z\right)  $ is an asymptotic series. \ However, as discussed
earlier in a general context, the $J\left(  \overline{w},z\right)  $ and
$K\left(  \overline{w},z\right)  $ series convergence properties can be
reversed by making compensating changes in the normalizations: \ $\left\{
J_{j},A_{k}\right\}  \rightarrow\left\{  Z_{j}J_{j},A_{k}/Z_{k}\right\}  $.
\ We will give other examples of kernels, in closed form, when we discuss the
corresponding quantum systems below.

\paragraph{Generalizations}

The $\left\{  J_{n}\left(  z\right)  ,A_{n}\left(  z\right)  \right\}  $
biorthogonal system has been generalized by Gegenbauer to other situations
involving Bessel functions with non-integral indices and their associated
polynomials (see \S 9.2 in \cite{Watson}). \ This is relevant to our
discussion, given below, of some simple magnetic field effects. \ Extensions
to confluent hypergeometric functions $\left.  _{1}F_{1}\right.  $ have been
proposed by Erd\'{e}lyi \cite{Erdelyi}. \ These are relevant to $H=\left(
p+\nu\right)  ^{2}+\mu_{1}\exp\left(  ix\right)  +\mu_{2}\exp\left(
2ix\right)  $\ for arbitrary $\mu_{1}$ and $\mu_{2}$, as also discussed below.
\ We will further generalize the polynomials to all systems governed by
Hamiltonians of the form (\ref{H}). \ But, before fleshing out the
biorthogonal structure in the various soluble quantum mechanical models of
interest to us, we give a survey of the classical dynamics for the first such
model. \ 

\section{Classical dynamics for $V\left(  x\right)  =\exp\left(  2ix\right)
$}

Consider the Hamiltonian%
\begin{equation}
H=p^{2}+m^{2}e^{2ix} \label{PTLiouvilleHamiltonian}%
\end{equation}
where $m$ is real. \ This is a complex form of the well-studied Liouville
dynamics \cite{BCGT,Nakayama}. \ The coefficient $m^{2}$ of the exponential
has to be real for the Hamiltonian to be PT symmetric. \ That is to say,
$H$\ is invariant under $p\rightarrow p$, $x\rightarrow-x$, and complex
conjugation. \ Or at least, that is the case when we restrict to \emph{real}
$x$ and \emph{real} $p$. \ These transformations are inspired by the
conventional PT transformations of quantum mechanics, for wave functions
$\psi\left(  x,t\right)  $ that depend on real $x$ and $t$. \ The phase of
$m^{2}$ can be changed to \emph{any} arbitrary value just by making a real
shift in $x$, but only shifts $\Delta x=n\pi/2$ for integer $n$ will preserve
PT invariance. \ Still, we may say all values of $m$, even if complex, give
theories translationally equivalent to PT symmetric models.

While this has to be the simplest periodic complex potential example one can
imagine, nevertheless, it is not obvious \emph{a priori} how or if the
dynamics is consistent, nor is it obvious what connections exist between a
classical system described by (\ref{PTLiouvilleHamiltonian}) and any quantum
deformation of that dynamics. \ A goal of this paper is to examine this
consistency in some detail, and to discuss these connections.

While simple, this is still an interesting dynamical system. \ However, there
is some ambiguity about what $H$ represents when $p$ and $x$ are complex.
\ Following Bender et al. \cite{BenderClassical}, we choose to continue the
model into the complex plane simply by complexification of the usual form of
Hamilton's equations of motion, without complex conjugation of any of the
variables appearing in those equations. \ Thus we take
(\ref{PTLiouvilleHamiltonian}) and
\begin{equation}
\frac{dx\left(  t\right)  }{dt}=\frac{\partial H}{\partial p}=2p\left(
t\right)  \ ,\ \ \ \frac{dp\left(  t\right)  }{dt}=-\frac{\partial H}{\partial
x}=-2im^{2}e^{2ix\left(  t\right)  } \label{Dynamics}%
\end{equation}
to hold for all complex $x$ and $p$. \ With this definition of the
continuation, the \emph{complex energy} is conserved $dH/dt=0$. \ So under
this complexification procedure there are in fact two conserved real
quantities, $\operatorname{Re}H$ and $\operatorname{Im}H$.

Complex energy conservation allows reduction to a single first order equation
\begin{equation}
\frac{dx\left(  t\right)  }{dt}=2p\left(  t\right)  =\pm2\sqrt{E-m^{2}%
e^{2ix\left(  t\right)  }} \label{ComplexRootEMinusV}%
\end{equation}
with integration in terms of elementary functions. \ This yields the classical
motion.%
\begin{equation}
x\left(  t\right)  =\frac{1}{2i}\ln\left(  \frac{E/m^{2}}{\cosh^{2}\left(
\mp2i\sqrt{E}t+\operatorname{arctanh}\frac{\sqrt{E-m^{2}e^{2ix\left(
0\right)  }}}{\sqrt{E}}\right)  }\right)  \label{ClassicalMotion}%
\end{equation}
Sign flips on the RHSs of (\ref{ComplexRootEMinusV}) and
(\ref{ClassicalMotion}) occur at \emph{complex turning points}, as given by
$\operatorname{Re}\sqrt{E-m^{2}e^{2ix\left(  t\right)  }}=0=\operatorname{Im}%
\sqrt{E-m^{2}e^{2ix\left(  t\right)  }}$. \ If no $\operatorname{Re}%
\sqrt{E-m^{2}e^{2ix\left(  t\right)  }}=0$ points are encountered, upon taking
the upper signs in (\ref{ComplexRootEMinusV}) and (\ref{ClassicalMotion}), the
motion is a continuous but varying progression towards positive
$\operatorname{Re}x$ with periodic oscillations in $\operatorname{Im}x$. \ We
plot a representative $E>0$ example ($E=1/4$, $x\left(  0\right)  =1$,
$\operatorname{Re}p\left(  0\right)  >0$) showing real and imaginary parts of
both $x$ and $p$ as parametric functions of $t$.

\hspace{-1.25in}%
%TCIMACRO{\FRAME{itbpFU}{8.5322in}{2.009in}{0in}{\Qcb{$\left(
%\operatorname{Re}x,\operatorname{Im}x\right)  $ in blue and $\left(
%\operatorname{Re}p,\operatorname{Im}p\right)  $ in red, plotted parametrically
%for $E=1/4$}}{}{biorthogonalqmforjmp__1.eps}%
%width 8.5322in;  height 2.009in;  depth 0in;  original-width 8.5167in;
%original-height 1.97in;  cropleft "0";  croptop "1";  cropright "0.9931";
%filename 'BiorthogonalQMforJMP__1.eps';file-properties "XNPEU";}}}%
%BeginExpansion
{\parbox[b]{8.5322in}{\begin{center}
\includegraphics[
trim=0.000000in 0.000000in 0.058765in 0.000000in,
height=2.009in,
width=8.5322in
]%
{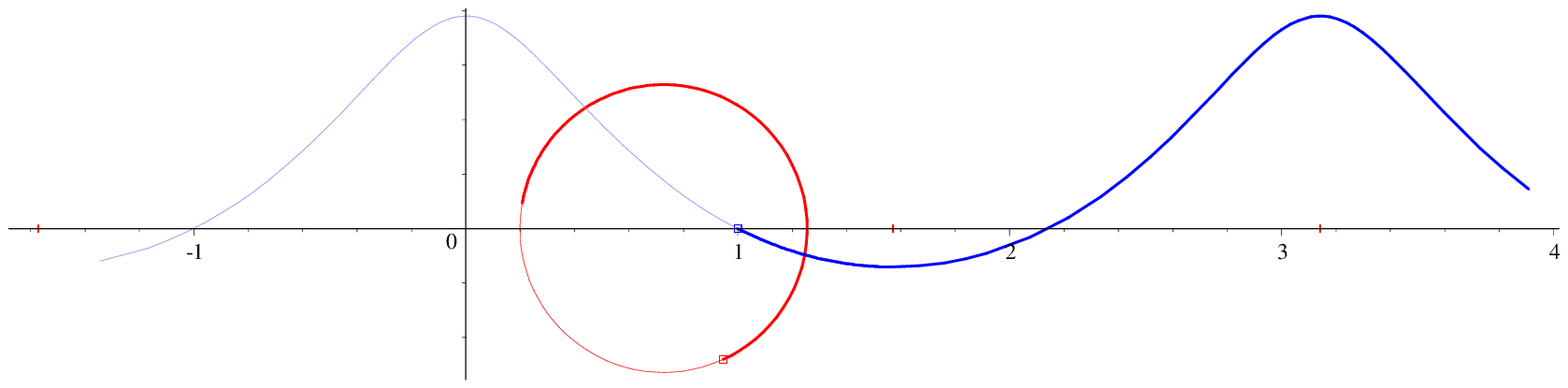}%
\\
$\left(  \operatorname{Re}x,\operatorname{Im}x\right)  $ in blue and $\left(
\operatorname{Re}p,\operatorname{Im}p\right)  $ in red, plotted parametrically
for $E=1/4$%
\end{center}}}%
%EndExpansion
\bigskip

\noindent In this and all other plots to follow, we have set $m=1$.
\ Heavier/lighter curves indicate forward/backward evolution in $t$. \ Initial
points are indicated by small boxes in the Figure.

The momentum trajectory in the Figure is as simple as it looks: \ A circle.
\ For all real $E>0$ and any initial $x\left(  0\right)  $ the momentum
$p\left(  t\right)  $ evolves around a complex circle centered on the real $p$
axis. \ This can be shown by inserting (\ref{ClassicalMotion}) into
(\ref{ComplexRootEMinusV}) and manipulating the result to obtain%
\begin{equation}
p\left(  t\right)  =\pm\sqrt{E}\left(  \frac{a-e^{\pm4i\sqrt{E}t}}%
{a+e^{\pm4i\sqrt{E}t}}\right)  \label{pClassical}%
\end{equation}
where we have defined the complex amplitude%
\begin{equation}
a\equiv\frac{E}{m^{2}}\left(  1+\sqrt{1-\frac{m^{2}}{E}e^{2ix\left(  0\right)
}}\right)  ^{2}e^{-2ix\left(  0\right)  } \label{a}%
\end{equation}
Hence $a$ depends on the initial data but not on $t$. \ In this form, $a$ and
$p$ have transparent high energy limits. \ From (\ref{pClassical}) it is
straightforward to establish by direct calculation for real $E>0$ that
$p\left(  t\right)  $ evolves around a circle in the complex plane, of radius
$\frac{2\left\vert a\right\vert }{\left\vert a\right\vert ^{2}-1}\sqrt{E}$,
whose center is on the real axis at $\pm\frac{\left\vert a\right\vert ^{2}%
+1}{\left\vert a\right\vert ^{2}-1}\sqrt{E}$. \ It also follows from
(\ref{pClassical}) that the momentum's angular velocity around the circle is
\emph{not} constant, but varies. \ In addition, for real $E>0$ a turning point
is encountered if and only if $\left\vert a\right\vert =1$, implying the $p$
circle has infinite radius and degenerates into a vertical straight line in
the complex plane, as described by purely imaginary momentum trajectories
$p\left(  t\right)  =-i\sqrt{E}\tan\left(  2\sqrt{E}t\mp\frac{1}{2}\arg
a\right)  $. \ Correspondingly, the $x$ trajectory for $\left\vert
a\right\vert =1$ is along a vertical line in the complex plane, with fixed
$\operatorname{Re}x$. \ For example, this can occur when $x\left(  0\right)
=N\pi$ and $E=m^{2}$, in which cases the $x$ trajectory \textquotedblleft
bounces off\textquotedblright\ the real axis at $t=0$.

The $E=0$ motion is somewhat simpler, but still has interesting structure, and
is given as follows.%
\begin{equation}
x_{E=0}\left(  t\right)  =i\ln\left(  e^{-ix\left(  0\right)  }\pm2mt\right)
\ ,\ \ \ \frac{dx_{E=0}\left(  t\right)  }{dt}=\frac{\pm2im}{e^{-ix\left(
0\right)  }\pm2mt}%
\end{equation}
Again, the momentum evolves along a circle of finite radius in the complex
plane, approaching the origin as a limit, except in the special situations
$x\left(  0\right)  =N\pi$, which correspond to circles of infinite radius.
\ In these special cases, the particle stays at $\operatorname{Re}x_{E=0}%
=N\pi$ for all $t$, but moves in the imaginary direction, with%
\begin{align}
\left.  \operatorname{Im}x_{E=0}\left(  t\right)  \right\vert
_{\operatorname{Re}x_{E=0}=N\pi}  &  =\frac{1}{2}\ln\left(  1+4m^{2}t^{2}%
\pm4mt\left(  -\right)  ^{N}\right) \nonumber\\
&  =\ln\left(  1\pm2mt\right)  \label{x(0)=NPi}%
\end{align}
and with $\pm$ signs depending on $N$ as well as the choice of root for
$p_{E=0}$. \ So with the $\pm$\ as exhibited in (\ref{x(0)=NPi}), $x_{E=0}$
goes to $x=+i\infty$ as $t\rightarrow\pm\infty$ and goes to $x=-i\infty$ as
$t\rightarrow\mp\frac{1}{2m}$. \ In fact, no matter what $x\left(  0\right)  $
is, the $E=0$ solution always has an imaginary part that behaves as
$\operatorname{Im}x_{E=0}\left(  t\right)  \ _{\widetilde{t\rightarrow
\pm\infty}}\ \ln\left(  t^{2}\right)  $ for at least one choice of sign. \ 

We plot parametrically some $E=0$ examples, for various initial $x\left(
0\right)  $, showing real and imaginary parts of $x\left(  t\right)  $ and
$p\left(  t\right)  $ for forward/backward evolution. \ The initial positions
are $x\left(  0\right)  =\pi/2$ (blue), $2.5$ (green), $2.8$ (black), and
$\pi$ (red), as indicated by small boxes in the Figure on the left.
\ Corresponding initial values for $p$ are also indicated by small boxes in
the Figure on the right.

\noindent%
%TCIMACRO{\FRAME{itbpFU}{3.2621in}{3.0018in}{0in}{\Qcb{$\left(
%\operatorname{Re}x,\operatorname{Im}x\right)  $ for $E=0$ and various initial
%positions}}{}{biorthogonalqmforjmp__2.eps}%
%width 3.2621in;  height 3.0018in;  depth 0in;  original-width 3.2508in;
%original-height 2.9827in;  cropleft "0";  croptop "0.9918";
%cropright "0.9895";  cropbottom "0";
%filename 'BiorthogonalQMforJMP__2.eps';file-properties "XNPEU";}}}%
%BeginExpansion
{\parbox[b]{3.2621in}{\begin{center}
\includegraphics[
trim=0.000000in 0.000000in 0.034133in 0.024458in,
height=3.0018in,
width=3.2621in
]%
{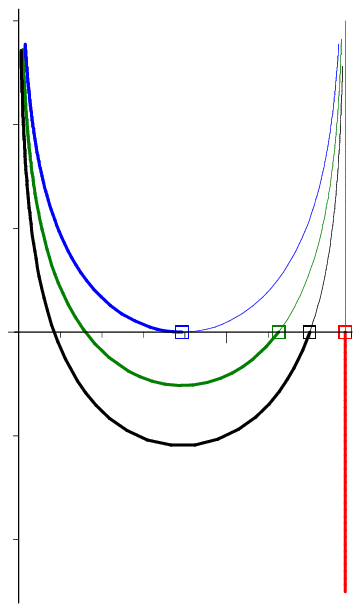}%
\\
$\left(  \operatorname{Re}x,\operatorname{Im}x\right)  $ for $E=0$ and various
initial positions
\end{center}}}%
%EndExpansion%
%TCIMACRO{\FRAME{itbpFU}{3.2621in}{3.0018in}{0in}{\Qcb{$\left(
%\operatorname{Re}p,\operatorname{Im}p\right)  $ for $E=0$ and various initial
%momenta}}{}{biorthogonalqmforjmp__3.eps}%
%width 3.2621in;  height 3.0018in;  depth 0in;  original-width 3.2508in;
%original-height 2.9827in;  cropleft "0";  croptop "0.9918";
%cropright "0.9895";  cropbottom "0";
%filename 'BiorthogonalQMforJMP__3.eps';file-properties "XNPEU";}}}%
%BeginExpansion
{\parbox[b]{3.2621in}{\begin{center}
\includegraphics[
trim=0.000000in 0.000000in 0.034133in 0.024458in,
height=3.0018in,
width=3.2621in
]%
{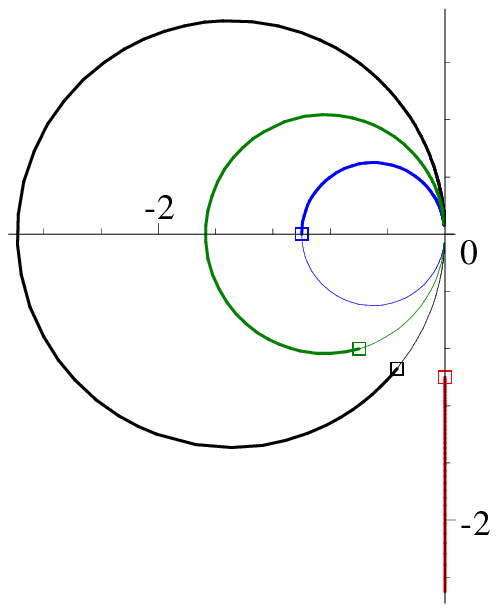}%
\\
$\left(  \operatorname{Re}p,\operatorname{Im}p\right)  $ for $E=0$ and various
initial momenta
\end{center}}}%
%EndExpansion

\paragraph{Canonical transformations}

Many of the classical properties are more easily understood upon taking note
of the following fact. \ There exist generating functions for canonical
transformations\ from Liouville to free particle dynamics \cite{BCT}, with
free variables $\theta$ and $p_{\theta}$. \ However, in the present context,
\emph{both} free particle and Liouville dynamics must be complexified, in
general. \ One such transformation is given by%
\begin{equation}
F=me^{ix}\sin\theta\ ,\ \ \ p\equiv\frac{\partial}{\partial x}F=ime^{ix}%
\sin\theta\ ,\ \ \ p_{\theta}\equiv-\frac{\partial}{\partial\theta}%
F=-me^{ix}\cos\theta\label{ClassicalTransformation}%
\end{equation}
The free particle's momentum is conserved, $p_{\theta}=\pm\sqrt{E}$, since
under the transformation%
\begin{equation}
H=p^{2}+m^{2}e^{2ix}=p_{\theta}^{2}=H_{\text{free}}%
\end{equation}
For real $E>0$, $p_{\theta}$ is also real. \ Although this would in general
require complex $x$, $p$, and $\theta$, as follows from the transformation.
\ Separating out the $\left(  x,p\right)  $ and $\left(  \theta,p_{\theta
}\right)  $ variables in (\ref{ClassicalTransformation}), and expressing one
set in terms of the other, leads to
\begin{align}
x  &  =-i\ln\left(  \frac{p_{\theta}}{-m\cos\theta}\right)
\ ,\ \ \ p=-ip_{\theta}\tan\theta\label{ClassicalXtoTheta}\\
\theta &  =\arcsin\left(  -ie^{-ix}p/m\right)  \ ,\ \ \ p_{\theta}%
=-me^{ix}\cos\left(  \arcsin\left(  -ie^{-ix}p/m\right)  \right) \nonumber
\end{align}
where some care is needed if cuts are encountered. \ These relations lead most
easily to the result that $p\left(  t\right)  $ lies on a complex circle, for
real $E>0$, as a simple consequence of $\theta\left(  t\right)  =\theta\left(
0\right)  +2p_{\theta}t$, where $\operatorname{Im}\theta\left(  t\right)
=\operatorname{Im}\theta\left(  0\right)  $ is an invariant for real $E>0$.
\ Comparing (\ref{ClassicalXtoTheta})\ to\ (\ref{pClassical}), we see that
$a=\exp\left(  -2i\theta\left(  0\right)  \right)  $, so $\left\vert
a\right\vert =\exp\left(  2\operatorname{Im}\theta\left(  0\right)  \right)
$. \ Thus, $x\left(  t\right)  $ turning points are encountered for real $E>0$
if and only if $\operatorname{Im}\theta=0$ on the corresponding free particle trajectories.

Actually, this canonical transformation works no matter what the energy is,
even if it is complex, in which case the classical trajectories can have an
elegant appearance. \ The Figure below shows a momentum trajectory for
$x\left(  0\right)  =1$ and $E=e^{i}$. \ The flow is between symmetrical
complex fixed points at $p_{\pm}=\pm e^{i/2}$.%
%TCIMACRO{\FRAME{dtbpF}{6.0226in}{3.0113in}{0pt}{}{}%
%{biorthogonalqmforjmp__4.eps}{\special{ language "Scientific Word";
%type "GRAPHIC";  maintain-aspect-ratio TRUE;  display "USEDEF";
%valid_file "F";  width 6.0226in;  height 3.0113in;  depth 0pt;
%original-width 6.0027in;  original-height 2.9922in;  cropleft "0";
%croptop "0.9916";  cropright "0.9934";  cropbottom "0";
%filename 'BiorthogonalQMforJMP__4.eps';file-properties "XNPEU";}}}%
%BeginExpansion
\begin{center}
\includegraphics[
trim=0.000000in 0.000000in 0.039618in 0.025135in,
height=3.0113in,
width=6.0226in
]%
{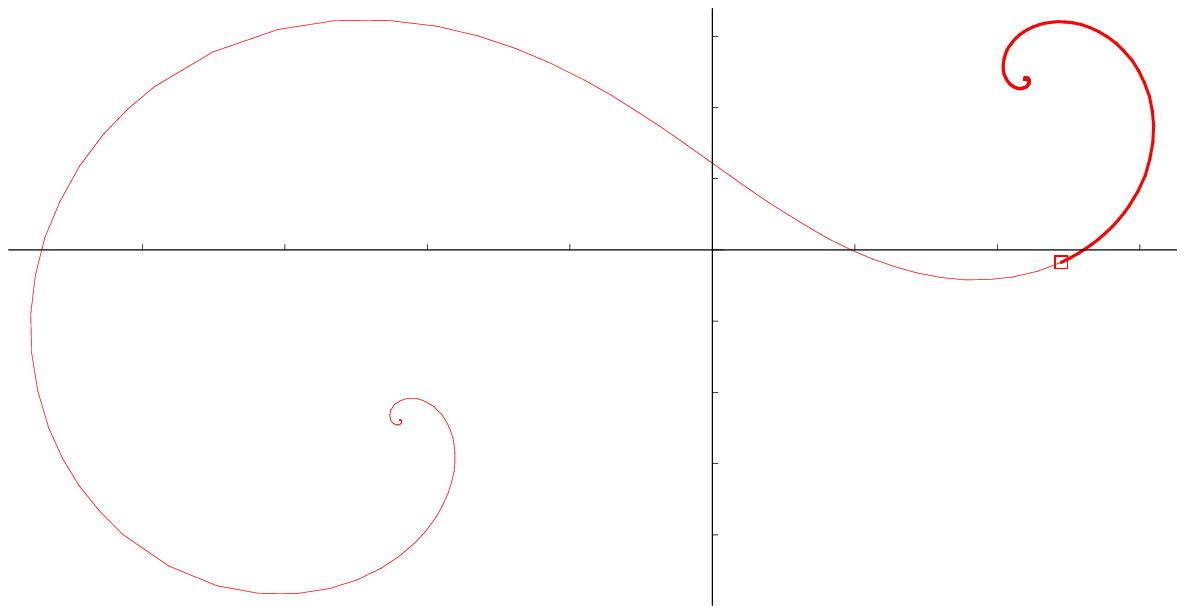}%
\end{center}
%EndExpansion
The choice $E=e^{2i}$ would give a more symmetrical trajectory, with $p\left(
t\right)  =-p\left(  -t\right)  $, and with a turning point. \ Of course, as
indicated by the explicit transformation (\ref{ClassicalXtoTheta}), complex
energy would require extending the free particle trajectories into both the
complex $\theta$ and the complex $p_{\theta}$ planes. \ 

The classical trajectories and the classical canonical transformation which we
have just described will appear explicitly in the quantum mechanics of the
model, to be discussed in the next section, where they play pivotal roles in
canonical integral representations of the wave functions and their duals.
\ But before ending our discussion of classical dynamics and turning our
attention to quantization, some comments on the canonical structure of this
system would seem to be in order. \ Indeed, this issue is brought into
particular focus by our previous remarks concerning classical canonical
transformations. \ 

We \emph{might} have proceeded as in the usual case of a particle on the
plane, and assumed the real and imaginary components of $x=\operatorname{Re}%
x+i\operatorname{Im}x$ and $p=\operatorname{Re}p+i\operatorname{Im}p$ are
independently canonical, up to some overall \textquotedblleft
isotropic\textquotedblright\ scale $\sigma$, with Poisson brackets (PBs) given
by $\left[  \operatorname{Re}x,\operatorname{Re}p\right]  _{\text{PB}}=\sigma
$,\ $\left[  \operatorname{Im}x,\operatorname{Im}p\right]  _{\text{PB}}%
=\sigma$,\ $\left[  \operatorname{Re}x,\operatorname{Im}x\right]  _{\text{PB}%
}=0$,\ $\left[  \operatorname{Re}x,\operatorname{Im}p\right]  _{\text{PB}}%
=0$,\ etc. \ Had we done so, we would then have had for \emph{any} value of
$\sigma$ the brackets $\left[  x,p\right]  _{\text{PB}}=0$,\ $\left[
x,H\right]  _{\text{PB}}=0$,\ $\left[  p,H\right]  _{\text{PB}}=0$. \ However,
this is \emph{not} the canonical structure that leads to Hamiltonian dynamics
as given in (\ref{Dynamics}), nor is it the structure that underlies the
classical canonical transformation given in (\ref{ClassicalTransformation}).
\ To obtain non-trivial equations of motion and other canonical
transformations of variables, in this isotropic Poisson bracket approach, it
would have been necessary to also conjugate some of the dynamical variables
appearing in $H$ and the generating function $F$.\ \ \ But we chose not to do this.

On the other hand, if we were to suppose the brackets are \emph{different }for
the real and imaginary parts,\ and modify them to be \emph{an}isotropic
brackets (ABs) as
\begin{equation}
\left[  \operatorname{Re}x,\operatorname{Re}p\right]  _{\text{AB}}=\sigma
+\cos^{2}\zeta\ ,\ \ \ \left[  \operatorname{Im}x,\operatorname{Im}p\right]
_{\text{AB}}=\sigma-\sin^{2}\zeta\label{ABs}%
\end{equation}
$\left[  \operatorname{Re}x,\operatorname{Im}x\right]  _{\text{AB}}%
=0$,\ $\left[  \operatorname{Re}x,\operatorname{Im}p\right]  _{\text{AB}}%
=0$,$\ $etc., then for \emph{any} values of $\sigma$ and $\zeta$ we would have%
\begin{equation}
\left[  x,p\right]  _{\text{AB}}=1\ ,\ \ \ \left[  x,H\right]  _{\text{AB}%
}=2p\ ,\ \ \ \left[  p,H\right]  _{\text{AB}}=-2im^{2}e^{2ix}
\label{ABconsequences}%
\end{equation}
This is exactly the dynamics specified in (\ref{Dynamics}) if we identify
$df\left(  x,p\right)  /dt=\left[  f\left(  x,p\right)  ,H\right]
_{\text{AB}}$, in the usual way. \ If we were to go further and impose the
additional condition that the bracket of $x$ with the complex conjugate of $p$
vanish, then we would be led to $2\sigma+\cos2\zeta=0$, hence $\sigma+\cos
^{2}\zeta=1/2$ and $\sigma-\sin^{2}\zeta=-1/2$. \ This would give the most
symmetrical form for the anisotropic brackets. \ For this choice the real and
imaginary components would have brackets differing by a naive PT
transformation, i.e. just a sign flip: $\left[  \operatorname{Re}%
x,\operatorname{Re}p\right]  _{\text{AB}}=1/2$,\ $\left[  \operatorname{Im}%
x,\operatorname{Im}p\right]  _{\text{AB}}=-1/2$. \ This is what we have done. \ 

We will briefly return to this issue at the end of the next section, in a
quantum phase-space framework, by considering deformations of the anisotropic brackets.

\section{Quantum mechanics for $V\left(  x\right)  =\exp\left(  2ix\right)  $}

Reconsider the PT symmetric Hamiltonian, (\ref{PTLiouvilleHamiltonian}), only
now in a quantum setting. \ Again, granted that PT\ symmetric quantum theories
make sense, this has to be the simplest periodic quantum potential problem one
can imagine.\footnote{This example is mentioned in passing by Bender et al.
\cite{BenderPeriodic}, without much discussion. \ It is considered in more
detail in \cite{Cannata}. \ Moreover, the model has been discussed extensively
in the mathematical literature, at least as early as the 1950s
\cite{Dunford,McGarvey,Rofe,Tkachenko,Gasymov,Veliev,Birnir,Pastur,Deift,Shin}%
. \ The model with periodic boundary conditions on $x\in\left[  0,2\pi\right]
$ is \emph{an exception} to the method used by Bender et al. to construct the
dual space. \ We will discuss this fully in the following.} \ The
corresponding Schr\"{o}dinger energy eigenvalue equation is%
\begin{equation}
\left(  -\frac{\partial^{2}}{\partial x^{2}}+m^{2}e^{2ix}\right)  \psi
_{E}=E\psi_{E} \label{Schrodinger}%
\end{equation}
Since the potential is periodic, with period $\pi$, Floquet's theorem assures
us that we can always find a solution of the form%
\begin{equation}
\psi\left(  x\right)  =e^{i\mu x}\phi\left(  x\right)
\end{equation}
where $\phi\left(  x+\pi\right)  =\phi\left(  x\right)  $. \ That is to say,
$\phi\left(  x\right)  $ has the form $\phi\left(  x\right)  =\sum
_{n\in\mathbb{Z}}c_{n}e^{2inx}$. \ These features are easily borne out, since
the equation can be solved exactly. \ Change variables to
\begin{equation}
z=me^{ix}%
\end{equation}
Then $\frac{\partial}{\partial x}=iz\frac{\partial}{\partial z}$ and
(\ref{Schrodinger}) becomes%
\begin{equation}
\left(  z^{2}\frac{\partial^{2}}{\partial z^{2}}+z\frac{\partial}{\partial
z}+z^{2}-E\right)  \psi\left(  z\right)  =0 \label{Bessel}%
\end{equation}
This of course is Bessel's equation, albeit in the perhaps unfamiliar
situation where the variable $z/m$ is on the unit circle. \ Series
solutions\ are explicitly given by%
\begin{equation}
J_{\pm\sqrt{E}}\left(  me^{ix}\right)  =\left(  \frac{m}{2}e^{ix}\right)
^{\pm\sqrt{E}}\sum_{n=0}^{\infty}\frac{\left(  -m^{2}/4\right)  ^{n}}%
{n!\Gamma\left(  1+n\pm\sqrt{E}\right)  }e^{2inx} \label{Solutions}%
\end{equation}
and exhibit features in accord with Floquet's theorem. \ For completeness, we
also recall the elegant contour integral representation of Schl\"{a}fli and
Sonine (\cite{Watson} \S 6.2) which is valid for \emph{all} $\sqrt{E}$ and
$x$.
\begin{equation}
J_{\pm\sqrt{E}}\left(  me^{ix}\right)  =\left(  \frac{m}{2}e^{ix}\right)
^{\pm\sqrt{E}}\frac{1}{2\pi i}\int_{-\infty}^{\left(  0+\right)  }w^{\mp
\sqrt{E}-1}\exp\left(  w-\frac{m^{2}e^{2ix}}{4w}\right)  dw \label{Sonine}%
\end{equation}
The contour begins at $-\infty$, with $\arg w=-\pi$, proceeds below the real
$w$\ axis towards the origin, loops in the positive, counterclockwise sense
around the origin (hence the $\left(  0+\right)  $ notation), and then
continues above the real $w$ axis back to $-\infty$, with $\arg w=+\pi$.

The two choices of sign on the index of the Bessel function give independent
solutions, so long as $\sqrt{E}$ is not an integer. \ The prefactor in
(\ref{Solutions})\ or (\ref{Sonine}) is just a free plane wave, for all real
$E\geq0$. \ The infinite sum (or Sonine's integral) provides \emph{periodic}
modulations and additional phases to the plane wave. \ Negative or complex $E
$ would give for the solutions either an exponential blow-up for $x$ on the
real line, or else non-periodic behavior for $x\in\left[  0,2\pi N\right]  $,
where in the latter case periodicity requires $\sqrt{E}=k/N$ for some integer
$k$. \ For these reasons negative or complex $E$ is always ruled to be
\emph{in}admissible. \ In this regard, we emphasize that our discussion
applies to $x$ on the real line, or else to $z=me^{ix}$ lying on a closed
contour surrounding the origin. \ For energy spectra corresponding to boundary
conditions on other contours in the complex plane, see \S 5 in \cite{Cannata}.

So, the solutions of (\ref{Schrodinger}) for all real $E\geq0$ are (recall
$J_{n}\left(  z\right)  =\left(  -1\right)  ^{n}J_{-n}\left(  z\right)  $ for
integer index) \
\begin{subequations}
\begin{align}
\psi_{E}^{+}\left(  x\right)  =\frac{1}{2}\left(  J_{\sqrt{E}}\left(
me^{ix}\right)  +J_{-\sqrt{E}}\left(  me^{ix}\right)  \right)   &
_{\overrightarrow{\sqrt{E}\rightarrow n\in\mathbb{N}}}\ \ \ \left\{
\begin{array}
[c]{cc}%
J_{n}\left(  me^{ix}\right)  & \text{for }n\text{ even}\\
0 & \text{for }n\text{ odd}%
\end{array}
\right. \\
\psi_{E}^{-}\left(  x\right)  =\frac{1}{2}\left(  J_{\sqrt{E}}\left(
me^{ix}\right)  -J_{-\sqrt{E}}\left(  me^{ix}\right)  \right)   &
_{\overrightarrow{\sqrt{E}\rightarrow n\in\mathbb{N}}}\ \ \ \left\{
\begin{array}
[c]{cc}%
0 & \text{for }n\text{ even}\\
J_{n}\left(  me^{ix}\right)  & \text{for }n\text{ odd}%
\end{array}
\right.
\end{align}
We have chosen linear combinations of $J_{\pm\sqrt{E}}$ for later convenience.
\ We have also chosen the phases of the eigenstates so that $\mathcal{PT\ }%
\psi_{E}^{\pm}\left(  x\right)  =\psi_{E}^{\pm}\left(  x\right)  $. \ When
$E=0$, since $\psi_{E=0}^{-}\equiv0$, we have a non-degenerate ground state,
$\psi_{E=0}^{+}\left(  x\right)  =J_{0}\left(  me^{ix}\right)  $. \ This is
unlike the case of \emph{real} Liouville theory, for which there is no ground
state \cite{DHoker}. \ 

In fact, note the lack of degeneracy whenever $\sqrt{E}$ is \emph{any}
integer. \ If $\sqrt{E}=n\geq0$, the two independent solutions of
(\ref{Bessel}) are usually taken to be $J_{\sqrt{E}}\left(  z\right)  $
and\ $\lim\limits_{\sqrt{E}\rightarrow n}Y_{\sqrt{E}}\left(  z\right)
=\lim\limits_{\sqrt{E}\rightarrow n}\frac{J_{\sqrt{E}}\left(  z\right)
\cos\left(  \pi\sqrt{E}\right)  -J_{-\sqrt{E}}\left(  z\right)  }{\sin\left(
\pi\sqrt{E}\right)  }$, the latter being well-defined even as $\sqrt
{E}\rightarrow$ integer. \ But, although $Y_{n}\left(  me^{ix}\right)  $ is
finite for real $x$, nevertheless it can still be ruled out as an admissible
solution because $Y_{n}$ involves a logarithm multiplying a periodic function.
\ Hence it has a periodically modulated term with an envelope linear in $x$,
so it is unbounded for $-\infty<x<\infty$. \ Alternatively, $Y_{n}\left(
me^{ix}\right)  $ is not a periodic function for $x\in\left[  0,2\pi\right]
$. \ So $Y_{n}\left(  me^{ix}\right)  $ is \emph{not} an admissible solution
on the domains of interest to us.

We plot $\left(  \operatorname{Re}\psi_{E}\left(  x\right)  ,\operatorname{Im}%
\psi_{E}\left(  x\right)  \right)  $ and $\left(  \operatorname{Re}\psi
_{E}^{2}\left(  x\right)  ,\operatorname{Im}\psi_{E}^{2}\left(  x\right)
\right)  $ parametrically, as functions of $x$, for two cases.%
%TCIMACRO{\FRAME{dtbpFU}{6.0226in}{4.0145in}{0pt}{\Qcb{$\left(
%\operatorname{Re}\psi_{E}\left(  x\right)  ,\operatorname{Im}\psi_{E}\right)
%$ and $\left(  \operatorname{Re}\psi_{E}^{2},\operatorname{Im}\psi_{E}%
%^{2}\right)  $, for $E=0$ (green and orange) and for $E=1/4$ (blue and red)}%
%}{}{biorthogonalqmforjmp__5.eps}{\special{ language "Scientific Word";
%type "GRAPHIC";  maintain-aspect-ratio TRUE;  display "USEDEF";
%valid_file "F";  width 6.0226in;  height 4.0145in;  depth 0pt;
%original-width 6.0027in;  original-height 3.9946in;  cropleft "0";
%croptop "0.9927";  cropright "0.9934";  cropbottom "0";
%filename 'BiorthogonalQMforJMP__5.eps';file-properties "XNPEU";}}}%
%BeginExpansion
\begin{center}
\includegraphics[
trim=0.000000in 0.000000in 0.039618in 0.029161in,
height=4.0145in,
width=6.0226in
]%
{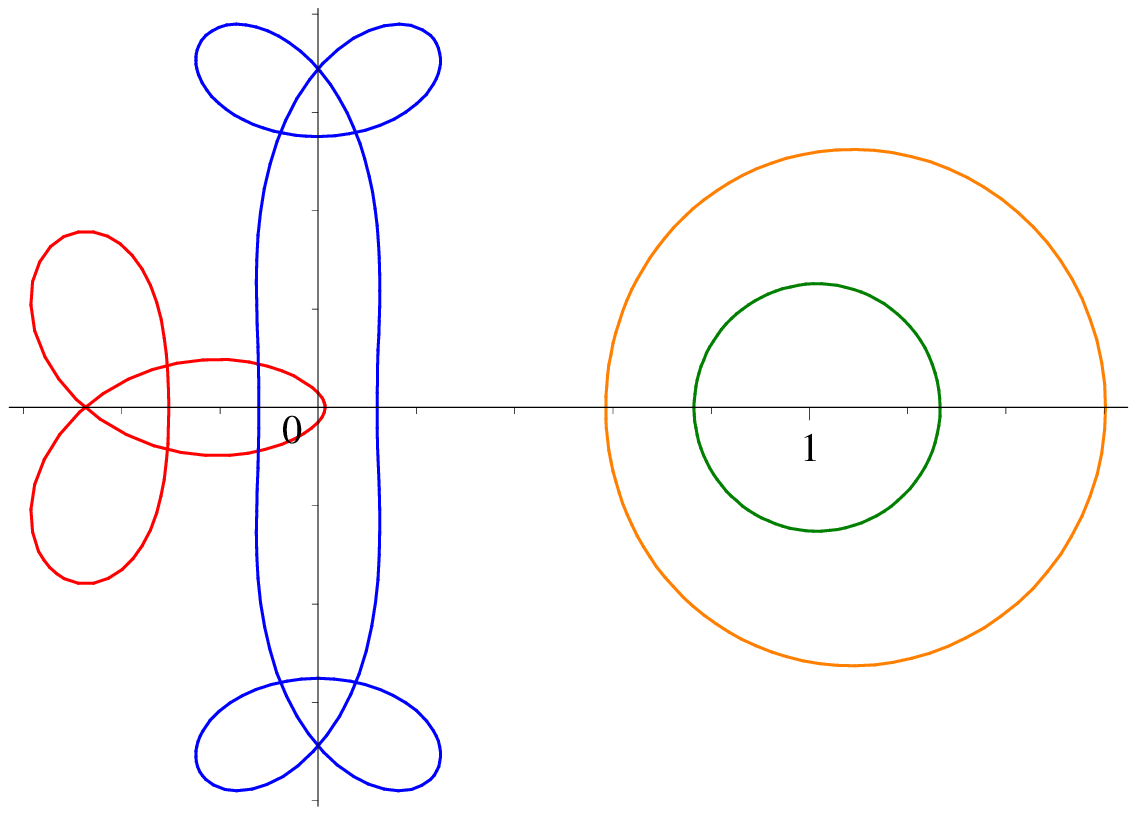}%
\\
$\left(  \operatorname{Re}\psi_{E}\left(  x\right)  ,\operatorname{Im}\psi
_{E}\right)  $ and $\left(  \operatorname{Re}\psi_{E}^{2},\operatorname{Im}%
\psi_{E}^{2}\right)  $, for $E=0$ (green and orange) and for $E=1/4$ (blue and
red)
\end{center}
%EndExpansion
\noindent In the case $E=0$, it is sufficient to take $x\in\left[
0,2\pi\right]  $. \ In the case $E=1/4$, we have plotted $\psi_{1/4}^{-}$ and
$\left(  \psi_{1/4}^{-}\right)  ^{2}$ for $x\in\left[  -2\pi,2\pi\right]  $.
\ The complete curve for the other wave function at this same energy,
$\psi_{1/4}^{+}$\ , is obtained by rotating the complete $\psi_{1/4}^{-}$
curve through $\pi/2$ about the origin, even though point-by-point,
$\psi_{1/4}^{+}\left(  x\right)  \neq e^{\pm i\pi/2}\psi_{1/4}^{-}\left(
x\right)  $. \ Similarly the complete $\left(  \psi_{1/4}^{+}\right)  ^{2}$
curve is obtained by rotating $\left(  \psi_{1/4}^{-}\right)  ^{2}$\ by $\pi.$ \ 

From these plots it is clear that $\psi^{2}\left(  x\right)  $ is not a bona
fide probability density since it is complex. \ It is also more or less clear
from the symmetry of $\psi^{2}$ that $\int_{0}^{2\pi}\left(  \psi_{0}\right)
^{2}dx$ is real and positive, while $\int_{-2\pi}^{2\pi}\left(  \psi_{1/4}%
^{-}\right)  ^{2}dx$ is real and negative. \ These are the so-called PT inner
products\ \cite{BenderReview}. \ This leads to the problem of constructing the
dual space for all the eigenfunctions.

When $\sqrt{E}$ is not an integer, the \textquotedblleft CPT\textquotedblright%
\ construction of the dual space\ \cite{BenderReview} goes through. \ This
follows because
\end{subequations}
\begin{equation}
\int_{-\infty}^{\infty}\psi_{E_{1}}^{\pm}\left(  x\right)  \psi_{E_{2}}^{\pm
}\left(  x\right)  dx=\pm\ \delta\left(  E_{1}-E_{2}\right)  \times
2\sin\left(  \pi\sqrt{E_{1}}\right)  \ ,\ \ \ \int_{-\infty}^{\infty}%
\psi_{E_{1}}^{+}\left(  x\right)  \psi_{E_{2}}^{-}\left(  x\right)  dx=0
\label{EigenfunctionOrthogonality}%
\end{equation}
and $\sin\left(  \pi\sqrt{E}\right)  $ on the RHS does not vanish when
$\sqrt{E}$ is not an integer. \ In this case, the dual functions can be
identified with
\begin{equation}
\chi_{E}^{\pm}=\frac{\pm1}{2\sin\left(  \pi\sqrt{E}\right)  }\psi_{E}^{\pm
}\left(  x\right)
\end{equation}
and exhibit the expected sign flips for alternating bands of energies.
\ However, when $\sqrt{E}$ is an integer, $\sin\left(  \pi\sqrt{E}\right)  $
vanishes, the PT symmetric eigenfunctions are self-orthogonal, and the CPT
construction of the dual space \emph{fails}. \ This happens for the invariant
subspace of non-degenerate eigenfunctions which are periodic on $x\in\left[
0,2\pi\right]  $. \ In the mathematical literature \cite{Gasymov,Veliev,Shin}
these integer $\sqrt{E}$ are known as spectral singularities\footnote{For
other examples of this effect in PT symmetric theories, see \cite{Samsonov}.},
a name seen to be appropriate from the behavior of the integrands for spectral
resolutions of projectors.%
\begin{equation}
\mathbb{P}_{\left[  E_{1},E_{2}\right]  }^{\pm}\left(  x,y\right)
=\int_{E_{1}}^{E_{2}}\psi_{E}^{\pm}\left(  x\right)  \psi_{E}^{\pm}\left(
y\right)  \frac{\pm1}{2\sin\left(  \pi\sqrt{E}\right)  }dE
\end{equation}
\emph{Henceforth we consider only the subspace of }$2\pi$\emph{-periodic}
\emph{eigenfunctions.}

\paragraph{A remarkable $2\pi$-periodic biorthogonal system}

Elements of the dual space for the $2\pi$-periodic eigenfunctions are given by
the Neumann polynomials as presented earlier. \ The key orthogonality relation
is now%
\begin{equation}
\frac{1}{2\pi}\int_{0}^{2\pi}A_{k}\left(  me^{ix}\right)  J_{n}\left(
me^{ix}\right)  dx=\delta_{kn} \label{AJOrthogonality}%
\end{equation}
where $A_{-n}\left(  z\right)  \equiv\left(  -1\right)  ^{n}A_{n}\left(
z\right)  $. \ This gives a remarkable biorthogonal system whose properties
warrant more discussion. \ The CPT\ method of constructing the dual space,
i.e. simply changing normalizations and phases of the PT transformed
functions, does \emph{not} provide a bi-orthonormalizable set of functions in
this case, since%
\begin{equation}
\frac{1}{2\pi}\int_{0}^{2\pi}J_{k}\left(  me^{ix}\right)  J_{n}\left(
me^{ix}\right)  dx=\delta_{k0}\delta_{n0}%
\end{equation}
This follows because the $J$s are series in only positive powers of $e^{ix}$.
\ So all the $2\pi$-periodic energy eigenfunctions are self-orthogonal except
for the ground state.

\paragraph{The effects of the Hamiltonian on the dual space}

The $A_{n}$ obey an inhomogeneous equation \cite{Watson}.%
\begin{equation}
-\frac{d^{2}}{dx^{2}}A_{n}\left(  me^{ix}\right)  +\left(  m^{2}e^{2ix}%
-n^{2}\right)  A_{n}\left(  me^{ix}\right)  =\left\{
\begin{array}
[c]{c}%
2nme^{ix}\text{ \ \ for }n\text{ odd}\\
2m^{2}e^{2ix}\text{ \ \ for }n\text{ even}%
\end{array}
\right.  \label{HActingOnDuals}%
\end{equation}
This leads to an interesting variant on the usual method of showing that
eigenfunctions and their duals are orthogonal for different energies.
\ Consider $\int_{0}^{2\pi}A_{k}\left(  me^{ix}\right)  HJ_{n}\left(
me^{ix}\right)  dx=n^{2}\int_{0}^{2\pi}A_{k}\left(  me^{ix}\right)
J_{n}\left(  me^{ix}\right)  dx$. \ Integrating by parts gives%
\begin{gather}
\int_{0}^{2\pi}A_{k}\left(  me^{ix}\right)  HJ_{n}\left(  me^{ix}\right)
dx=\int_{0}^{2\pi}J_{n}\left(  me^{ix}\right)  HA_{k}\left(  me^{ix}\right)
dx\nonumber\\
=k^{2}\int_{0}^{2\pi}J_{n}\left(  me^{ix}\right)  A_{k}\left(  me^{ix}\right)
dx+\left\{
\begin{array}
[c]{c}%
2km\int_{0}^{2\pi}e^{ix}J_{n}\left(  me^{ix}\right)  dx\text{ \ \ for }k\text{
odd}\\
\\
2m^{2}\int_{0}^{2\pi}e^{2ix}J_{n}\left(  me^{ix}\right)  dx\text{ \ \ for
}k\text{ even}%
\end{array}
\right.
\end{gather}
There are no boundary contributions upon integrating by parts since all
functions are $2\pi$-periodic. \ Moreover, the inhomogeneous terms are
orthogonal to the space of energy eigenfunctions. $\ $That is to say,
$\int_{0}^{2\pi}e^{ix}J_{n}\left(  me^{ix}\right)  dx=0=\int_{0}^{2\pi}%
e^{2ix}J_{n}\left(  me^{ix}\right)  dx$ since the $J$s are series in only
positive powers of $e^{ix}$. \ So even though (\ref{HActingOnDuals}) is
inhomogeneous, we still arrive at the usual conclusion.
\begin{equation}
\left(  n^{2}-k^{2}\right)  \int_{0}^{2\pi}A_{k}\left(  me^{ix}\right)
J_{n}\left(  me^{ix}\right)  dx=0
\end{equation}
hence $\int_{0}^{2\pi}A_{k}\left(  me^{ix}\right)  J_{n}\left(  me^{ix}%
\right)  dx=0$ if $k^{2}\neq n^{2}$. \ Recall that $A_{-n}\left(  z\right)
\equiv\left(  -1\right)  ^{n}A_{n}\left(  z\right)  $.

\paragraph{Integral representations for $E=n^{2}$ and quantum equivalence to a
free particle on a circle}

The $2\pi$-periodic Bessel functions are in fact the canonical integral
transforms of free plane waves on a circle, as constructed in this special
situation just by exponentiating the classical generating function presented
earlier in (\ref{ClassicalTransformation}). Explicitly,%
\begin{equation}
J_{n}\left(  me^{ix}\right)  =\frac{1}{2\pi}\int_{0}^{2\pi}e^{ime^{ix}%
\sin\theta}e^{-in\theta}d\theta\ ,\ \ \ n\in\mathbb{Z}
\label{CanonicalTransformationInteger}%
\end{equation}
The essence of the quantum equivalence lies in the identical action of $H$ and
$H_{\text{free}}=-\partial_{\theta}^{2}$ on the periodic kernel.%
\begin{equation}
He^{ime^{ix}\sin\theta}=H_{\text{free}}e^{ime^{ix}\sin\theta}=\left(
m^{2}e^{2ix}+\left(  m\frac{d}{dm}\right)  ^{2}\right)  e^{ime^{ix}\sin\theta}%
\end{equation}
While (\ref{CanonicalTransformationInteger}) is a well-known integral
representation of the Bessel function \cite{Watson}, the interpretation as a
canonical transformation is usually overlooked in most mathematical physics
texts. \ That is unfortunate. \ This is an extremely straightforward and
transparent connection between the quantum and classical theories that serves
to distinguish this model. \ (It doesn't get any better than
this!\footnote{The same situation applies to the real Liouville model. \ See
\cite{BCT,BCGT,Ghandour,Backlund,CurtrightGhandour,Cataplex}.}) \ 

The integral transform is a two-to-one map\ from the space of free particle
plane waves to Bessel functions ($e^{\mp in\theta}\rightarrow\left(
\pm1\right)  ^{n}J_{n}$ upon using $J_{-n}=\left(  -\right)  ^{n}J_{n}$), just
as happens in the real Liouville case ($e^{\pm ikx}\rightarrow K_{ik}$)
\cite{BCT,BCGT}, although here we have selected normalizations for the wave
functions so that the transformation uses the same normalization magnitude for
all $n$ (with at most only an $n$-dependent phase), unlike the real Liouville
integral tranformation, where the magnitude of the normalization varies with
$k$ in order to obtain a conventionally normalized continuum wave function
from the MacDonald function $K_{ik}$ (e.g. \cite{Cataplex}, 5\textit{th}
eqn.). \ 

When restricted to the combinations $e^{in\theta}+\left(  -\right)
^{n}e^{-in\theta}$ the map is one-to-one and invertible. \ On the other hand,
acting on $e^{in\theta}-\left(  -\right)  ^{n}e^{-in\theta}$ the map projects
to zero. \ When restricted either to the subspace of right-moving plane waves,
or to the subspace of left-moving plane waves, the map is also one-to-one, but
not invertible to the same subspace of plane waves. \ Inverting this
transformation on $J_{n}$ by using the \emph{same} kernel always gives a
combination of left- and right-movers. \ (See the discussion and Table below.)
\ This left-right combination is unavoidable, given the properties of the
kernel $\exp\left(  ime^{ix}\sin\theta\right)  $ and the symmetry
$J_{-n}=\left(  -\right)  ^{n}J_{n}$, although the combination of plane waves
\emph{can} be changed by choosing a different kernel in the integral
representation of $J_{n}$. \ (See the discussion following the Table.). \ In
any case, combinations of left- and right-movers result from inverting the map
using the kernel we have chosen.

We note the effects of index raising/lowering operations on the complete
integrand in (\ref{CanonicalTransformationInteger}).%
\begin{equation}
\left(  -i\frac{\partial}{\partial x}\mp n\right)  e^{ime^{ix}\sin
\theta-in\theta}=\left(  \mp i\frac{\partial}{\partial\theta}\mp me^{ix}e^{\mp
i\theta}\right)  e^{ime^{ix}\sin\theta-in\theta}%
\end{equation}
Thus, acting on the integral representation
(\ref{CanonicalTransformationInteger}) and integrating away the $\theta$
derivative term, we obtain the usual recursion relations \cite{Watson} only
expressed for Bessel functions on the circle.%
\begin{equation}
\left(  -i\frac{d}{dx}\mp n\right)  J_{n}\left(  me^{ix}\right)  =\mp
me^{ix}J_{n\pm1}\left(  me^{ix}\right)
\end{equation}
We also note in passing that the exponentiated classical generating function
has a well-known interpretation as an element of the Poincar\'{e} group on the
plane (\cite{Vilenkin} Chapter IV).

\paragraph{Inverting the transformations and the r\^{o}le of other classical
trajectories in the quantized model}

The canonical integral transformation from Neumann polynomials to plane waves
using the \emph{same} kernel as in (\ref{CanonicalTransformationInteger})
warrants discussion. \ It is simply%
\begin{equation}
\frac{1}{2\pi}\int_{0}^{2\pi}e^{ime^{ix}\sin\theta}A_{n}\left(  me^{ix}%
\right)  dx=\left\{
\begin{array}
[c]{ll}%
1 & \text{for }n=0\\
e^{in\theta}+\left(  -\right)  ^{n}e^{-in\theta} & \text{for }n>0
\end{array}
\right.  \label{AtoPlaneWave}%
\end{equation}
This follows from the biorthogonality of $\left\{  J_{j},A_{k}\right\}  $,
(\ref{AJOrthogonality}), and the well-known generating function (see
\cite{Abram} \textbf{9.1.41})%
\begin{equation}
e^{iz\sin\theta}=\sum_{n=-\infty}^{\infty}J_{n}\left(  z\right)  e^{in\theta
}=J_{0}\left(  z\right)  +\sum_{n=1}^{\infty}J_{n}\left(  z\right)  \left(
e^{in\theta}+\left(  -1\right)  ^{n}e^{-in\theta}\right)  \label{Generating}%
\end{equation}
as well as $J_{-n}\left(  z\right)  =\left(  -1\right)  ^{n}J_{n}\left(
z\right)  $. \ The transform (\ref{AJOrthogonality}) involves integration
around the circle, $z=me^{ix}$, $x\in\left[  0,2\pi\right]  $, of a Neumann
polynomial times the canonical transformation kernel. \ But, were we to make
slight deformations of the circular contour (as allowed by the analytic form
of the Neumann polynomials and the kernel in the neighborhood of the circle)
the result of the transformation would be the same. \ Therefore, the
transformation may be carried out just by integrating along a closed contour
in the $z=me^{ix}$ plane given exactly by following a single classical
Liouville particle $x$-trajectory, for $E>0$. \ 

To invert the transformation and go from plane waves to Neumann polynomials
using the same kernel ultimately requires taking free particle plane waves off
the circle and integrating up the imaginary axis. \ We may do this by first
integrating along an arc of the circle to the point $\theta=-\alpha$, and then
by following the path of a classical free particle trajectory for
$\operatorname{Im}p_{\theta}>0$, $\operatorname{Re}p_{\theta}=0$ which evolves
to $i\infty-\alpha$ as $t\rightarrow\infty$. \ Explicitly (\cite{Watson}
\S 9.14 Eqn(3))%
\begin{equation}
A_{n}\left(  z\right)  =\frac{1}{2}\varepsilon_{n}z\int_{0}^{\infty+i\alpha
}\left(  e^{n\vartheta}+\left(  -1\right)  ^{n}e^{-n\vartheta}\right)
e^{-z\sinh\vartheta}\cosh\vartheta d\vartheta\label{mystery}%
\end{equation}
where $\left\vert \alpha+\arg z\right\vert <\pi/2$. \ Integrating by parts
picks up a contribution at $0$ when $n$ is even, but not when $n$ is odd,
resulting in Schl\"{a}fli's version of the Neumann polynomials (see \S 9.3-34
in \cite{Watson}). \ Explicitly
\begin{equation}
S_{n}\left(  z\right)  \equiv\frac{2}{n}\left(  \frac{1}{\varepsilon_{n}}%
A_{n}\left(  z\right)  -\cos^{2}\left(  n\pi/2\right)  \right)  =\int
_{0}^{\infty+i\alpha}\left(  e^{n\vartheta}-\left(  -\right)  ^{n}%
e^{-n\vartheta}\right)  e^{-z\sinh\vartheta}d\vartheta
\end{equation}
Alternatively, in terms of the original free particle variable $\theta
=i\vartheta$.%
\begin{align}
S_{n}\left(  z\right)   &  =\frac{2}{n}\left(  \frac{1}{\varepsilon_{n}}%
A_{n}\left(  z\right)  -\cos^{2}\left(  n\pi/2\right)  \right)  =i\left(
-\right)  ^{n}\int_{0}^{i\infty-\alpha}\left(  e^{in\theta}-\left(  -\right)
^{n}e^{-in\theta}\right)  e^{iz\sin\theta}d\theta\nonumber\\
A_{n}\left(  z\right)   &  =\varepsilon_{n}\cos^{2}\left(  n\pi/2\right)
+i\varepsilon_{n}\left(  -1\right)  ^{n}\frac{n}{2}\int_{0}^{i\infty-\alpha
}\left(  e^{in\theta}-\left(  -\right)  ^{n}e^{-in\theta}\right)
e^{iz\sin\theta}d\theta\nonumber\\
&  =\left\{
\begin{array}
[c]{ll}%
\varepsilon_{n}-n\varepsilon_{n}\int_{0}^{i\infty-\alpha}e^{iz\sin\theta}%
\sin\left(  n\theta\right)  d\theta & \text{for }n\text{ even}\\
& \\
-in\varepsilon_{n}\int_{0}^{i\infty-\alpha}e^{iz\sin\theta}\cos\left(
n\theta\right)  d\theta & \text{for }n\text{ odd}%
\end{array}
\right.
\end{align}
At least three wedges (choices for $\alpha$) are required for an \emph{open}
cover of a full circle for $z$, although two wedges will cover all but two
points on the $z$-circle.

We give a summary Table of the canonical integral transforms using the kernel
$\exp\left(  ime^{ix}\sin\theta\right)  $. \ The paths of integration are
described in terms of relevant classical trajectories. \ In a well-defined
sense as indicated in the Table, only integrations over particular,
representative, classical trajectories are required to change wave functions
for the free particle into those for the Liouville system, and vice versa.
\ $\bigskip$

{\small \noindent\hspace{-0.55in}}$%
\begin{array}
[c]{ccc}%
\fbox{\textbf{Integral Representations}}\bigskip & \text{\textbf{with}} &
\fbox{\textbf{Relevant Trajectories}}\\
J_{n}\left(  z\right)  =\dfrac{1}{4\pi}%
%TCIMACRO{\dint _{0}^{2\pi}}%
%BeginExpansion
{\displaystyle\int_{0}^{2\pi}}
%EndExpansion
e^{iz\sin\theta}\left(  e^{-in\theta}+\left(  -1\right)  ^{n}e^{in\theta
}\right)  d\theta\bigskip & z\equiv me^{ix} &
\begin{array}
[c]{c}%
\theta\text{ on }\operatorname{Re}p_{\theta}>0\text{ , }\operatorname{Im}%
p_{\theta}=0\text{ }\\
\text{free particle trajectory}%
\end{array}
\\
\left\vert \cos\theta\right\vert
%TCIMACRO{\dint _{-\infty}^{\infty}}%
%BeginExpansion
{\displaystyle\int_{-\infty}^{\infty}}
%EndExpansion
e^{iz\sin\theta}J_{n}\left(  z\right)  dz=e^{in\theta}+\left(  -\right)
^{n}e^{-in\theta}\bigskip &
\begin{array}
[c]{c}%
\theta\text{ real, }\\
\sin\theta\neq\pm1\text{,}%
\end{array}
&
\begin{array}
[c]{c}%
x\text{ on }\operatorname{Re}E=0=\operatorname{Im}E\\
x\left(  0\right)  =0\text{ and }\pi\text{ }\\
\text{Liouville trajectories}%
\end{array}
\\
A_{n}\left(  z\right)  =\dfrac{\varepsilon_{n}z}{2i}%
%TCIMACRO{\dint _{0}^{i\infty-\alpha}}%
%BeginExpansion
{\displaystyle\int_{0}^{i\infty-\alpha}}
%EndExpansion
e^{iz\sin\theta}\left(  e^{-in\theta}+\left(  -1\right)  ^{n}e^{in\theta
}\right)  \cos\theta d\theta\bigskip &
\begin{array}
[c]{c}%
\left\vert \alpha+\arg z\right\vert <\pi/2\\
\text{i.e. }\left\vert \alpha+x\right\vert <\pi/2
\end{array}
&
\begin{array}
[c]{c}%
\theta\text{ on }\operatorname{Re}p_{\theta}<0\text{ , }\operatorname{Im}%
p_{\theta}=0\text{ , }\theta\left(  0\right)  =0\\
\text{and}\operatorname{Re}p_{\theta}=0\text{ , }\operatorname{Im}p_{\theta
}>0\text{ , }\theta\left(  0\right)  =\alpha\text{ }\\
\text{free particle trajectories}%
\end{array}
\\
\dfrac{1}{2\pi i}%
%TCIMACRO{\doint }%
%BeginExpansion
{\displaystyle\oint}
%EndExpansion
e^{iz\sin\theta}A_{n}\left(  z\right)  \dfrac{dz}{z}=\left(  e^{in\theta
}+\left(  -\right)  ^{n}e^{-in\theta}\right)  \dfrac{\varepsilon_{n}}%
{2}\bigskip & \varepsilon_{n}\equiv2-\delta_{n0} &
\begin{array}
[c]{c}%
x\text{ on }\operatorname{Re}E>0\text{ , }\operatorname{Im}E=0\\
\text{Liouville trajectory}%
\end{array}
\end{array}
$\bigskip

\noindent Analyticity is in control here, in the Table. \ Up to some clearly
discernible limitations, variations of the integration paths will not change
the results. \ Indeed, some deformation of the integration path may be
necessary to have it exactly coincide with the indicated classical trajectory
(for example, in the last case).

As previously remarked, the first and last rows of the Table are immediate
consequences of the biorthogonality of $\left\{  J_{j},A_{k}\right\}  $,
(\ref{AJOrthogonality}), the well-known generating function, (\ref{Generating}%
), and $J_{-n}\left(  z\right)  =\left(  -1\right)  ^{n}J_{n}\left(  z\right)
$ as well as $A_{-n}\left(  z\right)  \equiv\left(  -1\right)  ^{n}%
A_{n}\left(  z\right)  $. \ The second row follows from the first row and the
Fourier integral representation of the Dirac delta. \ This particular relation
is sometimes written an as integral representation for the Chebyshev
polynomials, after translating $\theta\rightarrow\theta+\pi/2$ and defining
$\omega\equiv\cos\theta$, whereupon it becomes (see \cite{Abram}
\textbf{11.4.24})\
\begin{equation}
\cos\left(  n\theta\right)  \equiv T_{n}\left(  \omega\right)  =\frac{1}%
{2}i^{n}\sqrt{1-\omega^{2}}\int_{-\infty}^{\infty}e^{-is\omega}J_{n}\left(
s\right)  ds\text{ \ \ for \ \ }\omega^{2}<1
\end{equation}
The third row of the Table is obtained just by substituting $\theta
=i\vartheta$ in (\ref{mystery}), with no further insight on our part, but it
may well have a more physical interpretation \cite{Gutzwiller}. \ 

For the kernel being used, as presented in the Table, the map is $\left\{
J_{n}\right\}  \leftrightarrow\left\{  e^{in\theta}+\left(  -\right)
^{n}e^{-in\theta}\right\}  \leftrightarrow\left\{  A_{n}\right\}  $. \ The
intermediate subspace of plane waves is not chiral. \ However, projections
onto chiral subspaces may also be constructed.

\paragraph{Chiral kernels}

The result (\ref{Generating}) is a bilinear but non-symmetric kernel that maps
linear combinations of plane waves onto integer Bessel functions.
\ Alternatively, the kernel maps Neumann polynomials onto equal mixtures of
left- and right-moving plane waves. \ But obviously, we may project onto half
the space of plane waves in an infinite number of other ways, say by choosing
to keep in the sum over $n$ only one of $\exp\left(  \pm in\theta\right)  $,
for all $n\geq0$, and thereby produce other maps that act nontrivially on the
complementary set of plane waves. \ 

For example, the left-moving chiral subspace of plane waves is mapped
one-to-one onto Bessel functions by the chiral kernel\footnote{The
right-moving chiral subspace can be similarly mapped onto Bessels using
$\widetilde{S}\left(  z,\theta\right)  =S\left(  z,-\theta\right)  $. \ Also
note the radius of the free particle circle can be modified, $e^{i\theta
}\rightarrow Me^{i\theta}$ as in $z=me^{ix}$, by giving $\theta$ a constant
imaginary part. \ }%
\begin{equation}
S\left(  z,\theta\right)  \equiv\sum_{n=0}^{\infty}J_{n}\left(  z\right)
e^{in\theta} \label{S}%
\end{equation}
Thus $S:\left\{  e^{-in\theta}\ |\ n\geq0\right\}  \rightarrow\left\{
J_{n}\left(  z\right)  \right\}  .$ \ Alternatively, all Neumann polynomials
are \textquotedblleft right-mapped\textquotedblright\ by the kernel into the
space of right-moving plane waves, \ $\left\{  A_{n}\left(  z\right)
\right\}  :S\rightarrow\left\{  e^{in\theta}\ |\ n\geq0\right\}  $.
\ Explicitly%
\begin{equation}
J_{n}\left(  z\right)  =\frac{1}{2\pi}\int_{0}^{2\pi}S\left(  z,\theta\right)
e^{-in\theta}d\theta\ ,\ \ \ e^{in\theta}=\frac{1}{2\pi i}%
%TCIMACRO{\doint }%
%BeginExpansion
{\displaystyle\oint}
%EndExpansion
\frac{dz}{z}A_{n}\left(  z\right)  S\left(  z,\theta\right)
\end{equation}
Moreover, note the null projective property on almost all of the right-moving
subspace, $S:\left\{  e^{in\theta}\ |\ n>0\right\}  \rightarrow\left\{
0\right\}  $, as well as on all Bessel functions except for $J_{0}$, $\left\{
J_{n}\left(  z\right)  \ |\ n>0\right\}  :S\rightarrow\left\{  0\right\}  $. \ 

The chiral kernel can be expressed as special cases of Lommel's functions of
two variables (\cite{Watson} \S 16.5-16.59) \
\begin{equation}
S\left(  z,\theta\right)  =U_{0}\left(  ie^{i\theta}z,z\right)  -iU_{1}\left(
ie^{i\theta}z,z\right)  \label{Lommel}%
\end{equation}
but regrettably, there does not seem to be an elementary closed-form for $S$.
\ From rewriting (\ref{Generating}) as $\sum_{n=-\infty}^{\infty}t^{n}%
J_{n}\left(  z\right)  =\exp\frac{1}{2}z\left(  t-\frac{1}{t}\right)  $,
however, an integral representation follows immediately.%
\begin{equation}
S\left(  z,\theta\right)  =\dfrac{1}{2\pi i}%
%TCIMACRO{\doint _{\left\vert t\right\vert >1}}%
%BeginExpansion
{\displaystyle\oint_{\left\vert t\right\vert >1}}
%EndExpansion
e^{\frac{1}{2}z\left(  t-\frac{1}{t}\right)  }\frac{1}{t-e^{i\theta}}dt
\label{ChiralKernelInt}%
\end{equation}
Here the $t$ contour is counterclockwise around the origin with $\left\vert
t\right\vert >1$.\footnote{If $\left\vert t\right\vert <1$ for the contour,
instead of $S$ it yields $1-\widetilde{S}\left(  z,\theta+\pi\right)  $.}
\ The simple appearance of (\ref{ChiralKernelInt}) is somewhat deceptive.

As in the case of the previous kernel, $e^{iz\sin\theta}$, this $S$
establishes a \emph{quantum} equivalence between complex Liouville and free
particle systems. \ Only now the equivalence is with free chiral particles on
a circle. \ Once again, this is directly evident from the identical action of
$H$ and $H_{\text{free}}=p_{\theta}^{2}$ on the kernel.%
\begin{equation}
HS\left(  z,\theta\right)  =H_{\text{free}}S\left(  z,\theta\right)
=\sum_{n=0}^{\infty}n^{2}J_{n}\left(  z\right)  e^{in\theta} \label{HS}%
\end{equation}
This may be checked against the partial differential equations satisfied by
the Lommel functions in (\ref{Lommel}), or the contour representation in
(\ref{ChiralKernelInt}). \ From either of those results, or by manipulating
the sum in (\ref{HS}), there follow other forms for this bilocal Hamiltonian
kernel acting on the two spaces.

The chiral kernel can be inverted on the space of Bessel functions by the
\emph{formal} kernel $S^{-1}\left(  \theta,z\right)  .$%
\begin{equation}
S^{-1}\left(  \theta,z\right)  \equiv\sum_{n=0}^{\infty}e^{-in\theta}%
A_{n}\left(  z\right)  \ ,\ \ \ e^{-in\theta}=\frac{1}{2\pi i}%
%TCIMACRO{\doint }%
%BeginExpansion
{\displaystyle\oint}
%EndExpansion
\frac{dz}{z}S^{-1}\left(  \theta,z\right)  J_{n}\left(  z\right)  \label{Sinv}%
\end{equation}
Thus $S^{-1}:\left\{  J_{n}\left(  z\right)  \right\}  \rightarrow\left\{
e^{-in\theta}\ |\ n\geq0\right\}  $, and therefore $S^{-1}S=1:\left\{
e^{-in\theta}\ |\ n\geq0\right\}  \rightarrow\left\{  e^{-in\theta}%
\ |\ n\geq0\right\}  $ and $SS^{-1}=1:\left\{  J_{n}\left(  z\right)
\right\}  \rightarrow\left\{  J_{n}\left(  z\right)  \right\}  $.
\ Alternatively, $\left\{  e^{in\theta}\ |\ n\geq0\right\}  :S^{-1}%
\rightarrow\left\{  A_{n}\left(  z\right)  \right\}  $ since formally%
\begin{equation}
A_{n}\left(  z\right)  =\frac{1}{2\pi}\int_{0}^{2\pi}e^{in\theta}S^{-1}\left(
\theta,z\right)  d\theta
\end{equation}
The Liouville and free Hamiltonians act differently on $S^{-1}\left(
\theta,z\right)  $ due to the inhomogeneous terms in (\ref{HActingOnDuals}).
\ Acting term-by-term and summing the formal series we find%
\begin{align}
\left(  H-H_{\text{free}}\right)  S^{-1}\left(  \theta,z\right)   &
=2zi\partial_{\theta}\sum_{\text{odd }n\text{ }>0}e^{-in\theta}+2z^{2}%
\sum_{\text{even }n\text{ }\geq0}e^{-in\theta}\nonumber\\
&  =\frac{2ze^{-i\theta}\left(  1+e^{-2i\theta}\right)  +2z^{2}\left(
1-e^{-2i\theta}\right)  }{\left(  1-e^{-2i\theta}\right)  ^{2}}
\label{HonSinv}%
\end{align}
This has singularities as a function of $\theta$, but these can be avoided in
any integrations by changing the radius of the free particle circle (i.e. by
taking $\operatorname{Im}\theta\neq0$.) \ The RHS of (\ref{HonSinv}) is also
manifestly orthogonal to the right sector upon integration $%
%TCIMACRO{\toint }%
%BeginExpansion
{\textstyle\oint}
%EndExpansion
\frac{dz}{z}$.

This formal inverse allows us to express the equivalence to a free chiral
particle in a somewhat heuristic, but very concise way.%
\begin{equation}
H_{\text{free chiral}}=S^{-1}HS \label{HFreeChiral}%
\end{equation}
Or more explicitly%
\begin{equation}
H_{\text{free chiral}}\left(  \theta^{\prime},\theta\right)  =\frac{1}{2\pi i}%
%TCIMACRO{\doint }%
%BeginExpansion
{\displaystyle\oint}
%EndExpansion
\frac{dz}{z}S^{-1}\left(  \theta^{\prime},z\right)  HS\left(  z,\theta\right)
=\sum_{n=0}^{\infty}e^{-in\theta^{\prime}}n^{2}e^{in\theta}=\frac{e^{i\left(
\theta-\theta^{\prime}\right)  }\left(  1+e^{i\left(  \theta-\theta^{\prime
}\right)  }\right)  }{\left(  1-e^{i\left(  \theta-\theta^{\prime}\right)
}\right)  ^{3}}%
\end{equation}
The last expression here is the usual form for the free chiral Hamiltonian
kernel, and it is manifestly \emph{not} local, whereas the complex $H$
\emph{is} local.

As a technical point, we note that $S^{-1}\left(  \theta,z\right)  $ is indeed
a formal kernel, since it is a divergent series, given the normalizations of
the Neumann polynomials. \ Nevertheless, it can be Borel summed in closed form
(a result due to Kapteyn, see \cite{Watson}, \S 9.16).%
\begin{equation}
\sum_{n=0}^{\infty}t^{n}A_{n}\left(  z\right)  =\left(  1+t^{2}\right)
z\int_{0}^{\infty}\frac{e^{-u}du}{\left(  1-t^{2}\right)  z-2tu}
\label{Kapteyn}%
\end{equation}
where the integral is recognizable as the exponential integral function
$\operatorname{Ei}$. \ Conversely, the sum in (\ref{Kapteyn}) is the
asymptotic expansion of the integral, the latter being well-defined and
obviously convergent when $\operatorname{Re}\left(  1-t^{2}\right)  z/t<0$.
\ As generally remarked following (\ref{DualBiNorm}), the convergence of this
generating function for the dual polynomials can be improved by making
compensating changes in the normalizations: \ $\left\{  J_{j},A_{k}\right\}
\rightarrow\left\{  Z_{j}J_{j},A_{k}/Z_{k}\right\}  $. Thus $Z_{n}$ can be
chosen to make $\sum_{n=0}^{\infty}\frac{1}{t^{n}}A_{n}\left(  z\right)
/Z_{n} $ absolutely, uniformly, and rapidly convergent. \ The price to be paid
is that $\sum_{n=0}^{\infty}Z_{n}J_{n}\left(  z\right)  e^{in\theta}$ will
then be a divergent series.\footnote{The most \textquotedblleft
balanced\textquotedblright\ asymptotic index behavior for rescaled
eigenfunctions and their duals is achieved as follows.%
\[
n!2^{n}J_{n}\left(  z\right)  \ _{\widetilde{n\rightarrow\infty}}%
\ z^{n}\ ,\ \ \ \frac{1}{n!2^{n}}A_{n}\left(  t\right)  \ _{\widetilde
{n\rightarrow\infty}}\ \frac{1}{t^{n}}%
\]
Then the leading behavior of the terms in the bilinear sums, modified
accordingly, is
\begin{align*}
\frac{1}{\left(  wz\right)  ^{N}}\sum_{n=N}^{\infty}\left(  n!2^{n}%
J_{n}\left(  w\right)  \right)  \left(  n!2^{n}J_{n}\left(  z\right)  \right)
\ _{\widetilde{N\rightarrow\infty}}\ \frac{1}{\left(  wz\right)  ^{N}}%
\sum_{n=N}^{\infty}w^{n}z^{n}  &  =\frac{1}{1-wz}\\
\left(  st\right)  ^{N}\sum_{n=N}^{\infty}\left(  \frac{1}{n!2^{n}}%
A_{n}\left(  s\right)  \right)  \left(  \frac{1}{n!2^{n}}A_{n}\left(
t\right)  \right)  \ _{\widetilde{N\rightarrow\infty}}\ \left(  st\right)
^{N}\sum_{n=N}^{\infty}\frac{1}{s^{n}t^{n}}  &  =\frac{1}{1-\frac{1}{st}}%
\end{align*}
}

Other non-symmetric kernels may also be constructed. \ For simple examples,
reconsider those in (\ref{CauchyIdent}) and (\ref{CauchyH}). \ Of greater
interest, perhaps, are symmetric kernels given as sums of bilinears where both
functions are of the same type. \ It is often possible to express these in
closed form as well.

\paragraph{A simple kernel for the bilocal norm}

While it may be difficult to find a closed form for the kernel $K\left(
x,y\right)  $, since our normalizations for the $A_{n}\left(  me^{ix}\right)
$\ make this a divergent series, it is a simple task to find a closed form for
$J\left(  x,y\right)  $, the inverse of $K$\ on the spaces of interest. \ We
recall Neumann's addition theorem, (\ref{AdditionTheorem}) and (\ref{AddThm}),
only this time we choose more convenient normalizations to write%
\begin{equation}
J\left(  x,y\right)  \equiv J_{0}\left(  me^{-ix}-me^{iy}\right)  =\sum
_{n=0}^{\infty}\varepsilon_{n}J_{n}\left(  me^{-ix}\right)  J_{n}\left(
me^{iy}\right)  \label{JKernel}%
\end{equation}
where $\varepsilon_{0}=1\ ,\ \varepsilon_{n\geq1}=2$. \ Formally, the
determinant of this kernel is%
\begin{align}
\det\left(  J\right)   &  =\exp\left(  Tr\ln J\right)  =\exp\int_{0}^{2\pi}%
\ln\left(  J\left(  x,x\right)  \right)  dx\nonumber\\
&  =\exp\int_{0}^{2\pi}\ln\left(  J_{0}\left(  2im\sin x\right)  \right)
dx\nonumber\\
&  =\exp\int_{0}^{2\pi}\ln\left(  I_{0}\left(  2m\sin x\right)  \right)  dx
\label{DetJ}%
\end{align}
where $I_{0}$ is the modified Bessel function (see (\ref{Imu}) below). \ Thus
$\det\left(  J\right)  $ never vanishes, for any real $m$, implying the
existence of the formal inverse for real $m$ and $x\in\left[  0,2\pi\right]
$. \ With the normalizations in (\ref{JKernel}) that formal inverse is
\begin{equation}
K\left(  x,y\right)  =\sum_{n=0}^{\infty}\frac{1}{\varepsilon_{n}}A_{n}\left(
me^{ix}\right)  A_{n}\left(  me^{-iy}\right)  \label{KKernel}%
\end{equation}
These exact results should be compared to those for the spectral resolvent, in
models where $V\left(  x\right)  =\left(  ix\right)  ^{N}$
\cite{Mezincescu,BenderExact,Dorey,Shin}. \ 

As in (\ref{DualBiNorm}), the manifestly hermitian kernel $J\left(
x,y\right)  $\ can be used to evaluate the norms of arbitrary $2\pi$-periodic
states
\begin{equation}
\psi\left(  x\right)  \equiv\sum_{n=0}^{\infty}c_{n}\sqrt{\varepsilon_{n}%
}J_{n}\left(  me^{ix}\right)  \label{Jpsi}%
\end{equation}
through use of the corresponding dual functions
\begin{equation}
\psi_{\text{dual}}\left(  x\right)  \equiv\sum_{n=0}^{\infty}c_{n}^{\ast}%
A_{n}\left(  me^{ix}\right)  /\sqrt{\varepsilon_{n}} \label{Adualpsi}%
\end{equation}%
\begin{equation}
\left\Vert \psi\right\Vert ^{2}=\frac{1}{\left(  2\pi\right)  ^{2}}\int
_{0}^{2\pi}dx\int_{0}^{2\pi}dy\overline{\ \psi_{\text{dual}}\left(  x\right)
}\ J\left(  x,y\right)  \ \psi_{\text{dual}}\left(  y\right)  =\sum
_{n=0}^{\infty}\left\vert c_{n}\right\vert ^{2} \label{JNorm}%
\end{equation}
We note that $J\left(  x,y\right)  $ converts the dual polynomials into
conjugated Bessel functions, and conjugated polynomials into
Bessels.\footnote{Simply changing the sign of $x$ in $J\left(  x,y\right)  $
yields a non-hermitian kernel that converts $A_{n}\left(  me^{iy}\right)  $
into non-conjugated $J_{n}\left(  me^{ix}\right)  $.}%
\begin{align}
J_{n}^{\ast}\left(  me^{ix}\right)   &  =J_{n}\left(  me^{-ix}\right)
=\frac{1}{2\pi\varepsilon_{n}}\int_{0}^{2\pi}dy\ J\left(  x,y\right)
A_{n}\left(  me^{iy}\right) \nonumber\\
J_{n}\left(  me^{iy}\right)   &  =\frac{1}{2\pi\varepsilon_{n}}\int_{0}^{2\pi
}dx\ A_{n}\left(  me^{-ix}\right)  J\left(  x,y\right)  \label{AtoJ*}%
\end{align}

\paragraph{Hamiltonian kernel on the dual space}

We also note that $J\left(  x,y\right)  $ converts $H$ into $H^{\ast}$ to
produce a bilocal Hamiltonian kernel equivalent to either, in the following
sense.%
\begin{equation}
\left(  -\frac{\partial^{2}}{\partial x^{2}}+m^{2}e^{-2ix}\right)  J\left(
x,y\right)  =\left(  -\frac{\partial^{2}}{\partial y^{2}}+m^{2}e^{2iy}\right)
J\left(  x,y\right)  =\sum_{n=0}^{\infty}n^{2}\varepsilon_{n}J_{n}\left(
me^{-ix}\right)  J_{n}\left(  me^{iy}\right)
\end{equation}
The sum converges absolutely, uniformly, and very rapidly. \ Evaluating this
sum in closed form produces an explicit hermitian kernel.
\begin{equation}
H\left(  x,y\right)  =H^{\ast}\left(  y,x\right)  =m\frac{J_{1}\left(
me^{-ix}-me^{iy}\right)  }{e^{-iy}-e^{ix}}=\sum_{n=0}^{\infty}n^{2}%
\varepsilon_{n}J_{n}\left(  me^{-ix}\right)  J_{n}\left(  me^{iy}\right)
\label{DualHamiltonianKernel}%
\end{equation}
As a series in $m\left(  e^{-ix}-e^{iy}\right)  $ this bilocal Hamiltonian is
rapidly convergent
\begin{equation}
H\left(  x,y\right)  =\frac{m^{2}}{2}e^{-i\left(  x-y\right)  }\sum
_{n=0}^{\infty}\frac{\left(  -m^{2}\right)  ^{n}\left(  e^{-ix}-e^{iy}\right)
^{2n}}{4^{n}n!\left(  n+1\right)  !}%
\end{equation}
and as a series in $e^{-ix}$ or $e^{iy}$, it only involves non-negative powers
of either. \ This Hamiltonian kernel acts on the dual space to give energy
operator matrix elements, or averages as in (\ref{fAverageBiLocal}).
\begin{equation}
\left\langle H\right\rangle _{\psi}=\frac{1}{\left(  2\pi\right)
^{2}\left\Vert \psi\right\Vert ^{2}}\int_{0}^{2\pi}dx\int_{0}^{2\pi
}dy\overline{\ \psi_{\text{dual}}\left(  x\right)  }\ H\left(  x,y\right)
\ \psi_{\text{dual}}\left(  y\right)  =\sum_{n=0}^{\infty}n^{2}\left\vert
c_{n}\right\vert ^{2}/\left\Vert \psi\right\Vert ^{2} \label{HAverage}%
\end{equation}
It is sometimes convenient to re-express this in terms of the action of
$H\left(  x,y\right)  $ on the dual polynomials, similar to (\ref{AtoJ*}).%
\begin{equation}
n^{2}J_{n}^{\ast}\left(  me^{ix}\right)  =n^{2}J_{n}\left(  me^{-ix}\right)
=\frac{1}{2\pi\varepsilon_{n}}\int_{0}^{2\pi}dy\ H\left(  x,y\right)
A_{n}\left(  me^{iy}\right)
\end{equation}

\paragraph{Another map between $H$ and $H^{\ast}$ on $2\pi$-periodic
functions}

Consider \cite{Cataplex}\ integral kernels depending on a parameter $s$. \ In
particular, consider the very simple kernel \
\begin{align}
K\left(  x,y;s\right)   &  \equiv\exp k\left(  x,y;s\right) \\
k\left(  x,y;s\right)   &  \equiv\frac{m}{2}\left(  \frac{1}{s}e^{i\left(
-x+y\right)  }-se^{i\left(  -x-y\right)  }-se^{i\left(  x+y\right)  }\right)
\nonumber\\
k\left(  x,y;s\right)   &  =k^{\ast}\left(  y,x;s\right)  \text{
\ \ for\ }s\in\mathbb{R}\nonumber
\end{align}
All terms in these expressions are periodic in $x$ and $y$, separately, with
period $2\pi$. \ Once again we have%
\begin{equation}
\left(  -\frac{\partial^{2}}{\partial x^{2}}+m^{2}e^{-2ix}\right)  K\left(
x,y;s\right)  =\left(  -\frac{\partial^{2}}{\partial y^{2}}+m^{2}%
e^{2iy}\right)  K\left(  x,y;s\right)  \label{SimpleKernel}%
\end{equation}
So this kernel also accomplishes the task of converting $H$ into $H^{\ast}$
and vice versa. \ Using this kernel, we find the integral
transforms\footnote{A similar kernel may be used in real Liouville theory to
effect an integral transformation from one $K_{\nu}$ to another, leading to
MacDonald's identity (\cite{Watson} \S 13.71). \ See \cite{Cataplex}\ and
references therein. \ We were not able to find (\ref{JtoJ*}) in the
literature. \ But it is not too difficult to work it out.}%
\begin{align}
J_{n}^{\ast}\left(  me^{ix}\right)   &  =J_{n}\left(  me^{-ix}\right)
=\frac{1}{2\pi}\frac{\left(  -1\right)  ^{n}}{I_{n}\left(  ms\right)  }%
\int_{0}^{2\pi}K\left(  x,y;s\right)  J_{n}\left(  me^{iy}\right)
dy\nonumber\\
J_{n}\left(  me^{iy}\right)   &  =\frac{1}{2\pi}\frac{\left(  -1\right)  ^{n}%
}{I_{n}\left(  ms\right)  }\int_{0}^{2\pi}J_{n}\left(  me^{-ix}\right)
K\left(  x,y;s\right)  dx \label{JtoJ*}%
\end{align}
which should be compared to (\ref{AtoJ*}). \ For real $ms\neq0$, the modified
Bessel $I_{n}\left(  ms\right)  $ is real and it has no zeroes, as is evident
in%
\begin{equation}
I_{\mu}\left(  ms\right)  =\left(  \frac{ms}{2}\right)  ^{\mu}\sum
_{k=0}^{\infty}\frac{1}{k!\Gamma\left(  1+\mu+k\right)  }\left(  \frac{ms}%
{2}\right)  ^{2k} \label{Imu}%
\end{equation}
For all real $m$ and $s$, $K\left(  x,y;s\right)  $ is an hermitian kernel,
$K\left(  x,y;s\right)  =K\left(  y,x;s\right)  ^{\ast}$, and it provides an
explicit, one-to-one, invertible, closed-form transformation between the
eigenfunctions of $H$ and $H^{\ast}$.

Another way to state the result (\ref{JtoJ*}) is as an expansion of the kernel
in terms of Bessel/Neumann trilinears. \ (A similar relation holds for real
Liouville theory. \ See \cite{Cataplex}.) \ From the transforms given, and the
completeness of the $J_{n}\left(  z\right)  $ on positive powers of $z$, we
deduce that%
\begin{align}
K\left(  x,y;s\right)   &  =J_{0}\left(  me^{-ix}\right)  I_{0}\left(
ms\right)  J_{0}\left(  me^{iy}\right) \nonumber\\
&  +\sum_{n=1}^{\infty}\left(  -1\right)  ^{n}J_{n}\left(  me^{-ix}\right)
I_{n}\left(  ms\right)  A_{n}\left(  me^{iy}\right)  +\left(  -1\right)
^{n}A_{n}\left(  me^{-ix}\right)  I_{n}\left(  ms\right)  J_{n}\left(
me^{iy}\right) \nonumber\\
&  +\sum_{j,k=1}^{\infty}J_{j}\left(  me^{-ix}\right)  c_{jk}\left(  s\right)
J_{k}\left(  me^{iy}\right)
\end{align}
where we have not determined the coefficients $c_{jk}\left(  s\right)  $
(except for the omitted $c_{00}=0=c_{01}=c_{10}$). \ 

More information may be incorporated into the sums expressing $K\left(
x,y;s\right)  $ by imposing total symmetry among $e^{-ix},e^{iy},$ and
$is\equiv e^{iz}$. \ Since $J_{n}\left(  me^{iz}=ims\right)  =i^{n}%
I_{n}\left(  ms=-ime^{iz}\right)  $ we have
\begin{subequations}
\begin{gather}
J_{n}\left(  me^{ix}\right)  J_{n}\left(  me^{iy}\right)  =\frac{1}{2\pi
i^{n}}\int_{0}^{2\pi}S\left(  x,y,z\right)  J_{n}\left(  me^{iz}\right)
dz\label{JtoJ}\\
S\left(  x,y,z\right)  \equiv\exp\left[  \frac{im}{2}\left(  e^{i\left(
x+y-z\right)  }+e^{i\left(  x-y+z\right)  }+e^{i\left(  -x+y+z\right)
}\right)  \right]  =K\left(  -x,y;s=-ie^{iz}\right)
\end{gather}
where $S$ is unchanged by permutations of its arguments. \ It follows as above
that this symmetric kernel is of the form
\end{subequations}
\begin{gather}
S\left(  x,y,z\right)  =J_{0}\left(  me^{ix}\right)  J_{0}\left(
me^{iy}\right)  J_{0}\left(  me^{iz}\right)  +\sum_{j,k,n=1}^{\infty}%
c_{jkn}\ J_{j}\left(  me^{ix}\right)  J_{k}\left(  me^{iy}\right)
J_{n}\left(  me^{iz}\right) \\
+\sum_{n=1}^{\infty}i^{n}A_{n}\left(  me^{ix}\right)  J_{n}\left(
me^{iy}\right)  J_{n}\left(  me^{iz}\right)  +i^{n}J_{n}\left(  me^{ix}%
\right)  A_{n}\left(  me^{iy}\right)  J_{n}\left(  me^{iz}\right)  +i^{n}%
J_{n}\left(  me^{ix}\right)  J_{n}\left(  me^{iy}\right)  A_{n}\left(
me^{iz}\right) \nonumber
\end{gather}
We have not determined the symmetric coefficients $c_{jkn}$. \ Regardless of
their values, they drop out of (\ref{JtoJ}).\ Their relation to the previous
coefficients is
\begin{equation}
c_{jk}\left(  s\right)  =\delta_{jk}i^{k}A_{k}\left(  ims\right)  +\sum
_{n=1}^{\infty}c_{jkn}i^{n}I_{n}\left(  ms\right)
\end{equation}
The $c_{jkn}$ could be determined by evaluating $\int_{0}^{2\pi}S\left(
x,y,z\right)  A_{n}\left(  me^{iz}\right)  dz$ through a direct calculation. \ 

\paragraph{Classical behavior of time-dependent expectation values}

As a final comment on the structure of this model, we remark that it is
straightforward to take linear combinations of energy eigenstates to produce
near-classical time-dependent behavior for expectation values, including those
which may be complex, such as
\begin{align}
\left\langle p\right\rangle _{\psi}\left(  t\right)   &  \equiv\frac{1}%
{2\pi\left\Vert \psi\right\Vert ^{2}}\int_{0}^{2\pi}dy\ \psi_{\text{dual}%
}\left(  y;t\right)  \left(  -i\partial_{y}\psi\left(  y;t\right)  \right) \\
&  =\frac{1}{\left(  2\pi\right)  ^{2}\left\Vert \psi\right\Vert ^{2}}\int
_{0}^{2\pi}dx\int_{0}^{2\pi}dy\ \overline{\psi_{\text{dual}}\left(
x;t\right)  }\left(  -i\partial_{y}J\left(  x,y\right)  \right)
\psi_{\text{dual}}\left(  y;t\right)
\end{align}
Here we have defined the time-dependent function and its dual as expected from
(\ref{Jpsi}) and (\ref{Adualpsi}).%
\begin{equation}
\psi\left(  x;t\right)  =\sum_{n=0}^{\infty}c_{n}e^{-in^{2}t}\sqrt
{\varepsilon_{n}}J_{n}\left(  me^{ix}\right)  \ ,\ \ \ \psi_{\text{dual}%
}\left(  x;t\right)  =\sum_{n=0}^{\infty}c_{n}^{\ast}e^{in^{2}t}A_{n}\left(
me^{ix}\right)  /\sqrt{\varepsilon_{n}} \label{TimeDependentPhases}%
\end{equation}
We have also used $\psi\left(  y;t\right)  =\frac{1}{2\pi}\int_{0}^{2\pi
}dx\ \overline{\psi_{\text{dual}}\left(  x;t\right)  }J\left(  x,y\right)  $
to write the expectation value in terms of the non-hermitian bilocal kernel
$-i\partial_{y}J\left(  x,y\right)  $.

In fact, a simple combination of only two energy eigenstates will give
$\left\langle p\right\rangle _{\psi}\left(  t\right)  $ that evolves around a
circle in the complex plane, whose center is on the real axis, with
\emph{constant} angular velocity around the circle. \ We believe that only
slightly more complicated coherent states \cite{Klauder} can be constructed to
reproduce the exact time-dependence evident in the classical motions, where
the angular velocity around the momentum circle is \emph{not} constant. \ All
this is in accord with the time derivative%
\begin{align}
\frac{d}{dt}\left\langle p\right\rangle _{\psi}\left(  t\right)   &
=-i\left\langle \left[  p,H\right]  \right\rangle _{\psi}\left(  t\right)
\equiv\frac{-i}{2\pi\left\Vert \psi\right\Vert ^{2}}\int_{0}^{2\pi}%
dx\ \psi_{\text{dual}}\left(  x;t\right)  \left[  p,H\right]  \psi\left(
x;t\right) \nonumber\\
&  =-2im^{2}\left\langle e^{2ix}\right\rangle _{\psi}\left(  t\right)
\label{d<p>/dt}%
\end{align}
Classical evolution of $\left\langle p\right\rangle _{\psi}\left(  t\right)  $
follows if the state $\psi$ is such that the last RHS evolves classically.
Since we also have%
\begin{equation}
\frac{d}{dt}\left\langle e^{2ix}\right\rangle _{\psi}\left(  t\right)
=-i\left\langle \left[  e^{2ix},H\right]  \right\rangle _{\psi}\left(
t\right)  =2i\left\langle \left\{  p,e^{2ix}\right\}  \right\rangle _{\psi
}\left(  t\right)  \label{d<exp(2ix)>/dt}%
\end{equation}
classical behavior is guaranteed to the extent that $\left\langle \left\{
p,e^{2ix}\right\}  \right\rangle _{\psi}\left(  t\right)  $ coincides with
$2\left\langle p\right\rangle _{\psi}\left(  t\right)  \left\langle
e^{2ix}\right\rangle _{\psi}\left(  t\right)  $. \ (This is just Ehrenfest's
theorem for this biorthogonal system.) \ In this regard, it is crucial that we
are able to replace $n^{2}A_{n}$ by $HA_{n}$ under the integrations over $x$
in (\ref{d<p>/dt}) and (\ref{d<exp(2ix)>/dt}). \ Fortunately, the
inhomogeneous terms in (\ref{HActingOnDuals}) do not contribute to the
expectations under consideration, so the replacement is permitted among the
integrands in these cases.

Remnants of classical behavior or the classical limits of quantum systems are
often most clearly understood from the point of view of phase-space quantum
mechanics, i.e. deformation quantization \cite{ZFC}. \ In that framework, let
us briefly return to the issue of the anisotropic brackets. \ The classical
anisotropic brackets described earlier are just $\hbar\rightarrow0$ limits of
Moyal brackets built from anisotropic Groenewold products acting on
phase-space distributions, where the phase-space variables are allowed to be
complex. \ Thus%
\begin{equation}
\left[  f\left(  x,p\right)  ,g\left(  x,p\right)  \right]  _{\text{AB}}%
=\lim_{\hbar\rightarrow0}\frac{1}{i\hbar}\left(  f\divideontimes
g-g\divideontimes f\right)  \label{AnisotropicMoyalBracket}%
\end{equation}
where the associative but non-commutative anisotropic product is given in its
most symmetrical form by%
\begin{equation}
\divideontimes\equiv e^{i\frac{\hbar}{4}\left(  \overleftarrow{\partial
}_{\operatorname{Re}x}\overrightarrow{\partial}_{\operatorname{Re}%
p}-\overleftarrow{\partial}_{\operatorname{Re}p}\overrightarrow{\partial
}_{\operatorname{Re}x}\right)  -i\frac{\hbar}{4}\left(  \overleftarrow
{\partial}_{\operatorname{Im}x}\overrightarrow{\partial}_{\operatorname{Im}%
p}-\overleftarrow{\partial}_{\operatorname{Im}p}\overrightarrow{\partial
}_{\operatorname{Im}x}\right)  } \label{AnisotropicGroenewold}%
\end{equation}
Clearly then, $x\divideontimes p=xp+i\hbar/2,\ p\divideontimes x=xp-i\hbar/2,
$ and $\left[  x,p\right]  _{\text{AB}}=1$ as in (\ref{ABconsequences}), as
well as $\left[  \operatorname{Re}x,\operatorname{Re}p\right]  _{\text{AB}%
}=1/2$,\ $\left[  \operatorname{Im}x,\operatorname{Im}p\right]  _{\text{AB}%
}=-1/2$, etc. \ Therefore the previous continuation of the classical dynamics
is recovered in the $\hbar\rightarrow0$ limit. \ In fact, the quantization of
the system can be carried out in full detail on the complex phase-space
through the use of this star product acting on Wigner functions built from
bilinears in the wave functions and their duals. \ (A thorough discussion of
this approach to quantization is in preparation \cite{CMV}.)

\section{Magnetic field effects}

Magnetic fields can produce some remarkable changes in the structure of a
complex biorthogonal system, even in the simple physical set-up where charged
particles move around a solenoid containing magnetic flux $\Phi$ with only a
constant vector potential $A\propto\Phi$ on the surface of the solenoid, in
addition to the complex potential energy imagined to be already
present.\footnote{Alternatively, non-hermitian $H$ could be obtained\ just by
considering complex vector potentials for a particle on the plane, without any
exponential interactions. \ See \cite{Fairlie}.} \ As an example, we add such
magnetic effects minimally to the previous Hamiltonian, to obtain the form
($\nu\equiv-eA/c$)%
\begin{equation}
H=\left(  p+\nu\right)  ^{2}+m^{2}e^{2ix}%
\end{equation}
Here $p$ is the momentum canonically conjugate to $x$, so classically there is
no discernible change from the case $\nu=0$. \ Classically, we have
$dx/dt=\left[  x,H\right]  =2\left(  p+\nu\right)  $, so $H=\frac{1}{4}\left(
dx/dt\right)  ^{2}+m^{2}e^{2ix}$, and for a given energy the trajectories are
the same as obtained in the $\nu=0$ case. \ A corresponding classical
canonical transformation taking $H\rightarrow H_{\text{free}}=p_{\theta}^{2} $
is now given by%
\begin{equation}
F=-\nu x+me^{ix}\sin\theta\ ,\ \ \ p\equiv\frac{\partial}{\partial x}%
F=-\nu+ime^{ix}\sin\theta\ ,\ \ \ p_{\theta}\equiv-\frac{\partial}%
{\partial\theta}F=-me^{ix}\cos\theta
\end{equation}
Thus there are no significant changes, classically.

But consider the QM of this model, and consider only $2\pi$-periodic
solutions, since the effects of $\nu\neq0$ are most dramatic in this
situation. \ As an operator%
\begin{equation}
H=\left(  -i\partial_{x}+\nu\right)  ^{2}+m^{2}e^{2ix} \label{Hnu}%
\end{equation}
and for generic $\nu$ the set of periodic eigenfunctions now consists of two
sectors given by%
\begin{equation}
H\ e^{-i\nu x}J_{\nu\pm n}\left(  me^{ix}\right)  =\left(  \nu\pm n\right)
^{2}e^{-i\nu x}J_{\nu\pm n}\left(  me^{ix}\right)
\end{equation}
for $n\in\mathbb{Z}_{\geq0}$. \ These eigenfunctions are all even under PT,
for real $\nu$. \ For want of better terms, we will call these the
\emph{right} and \emph{left sectors} since for generic $\nu$ all right-moving
and left-moving plane waves\ are contained in the spans of $\left\{  e^{-i\nu
x}J_{\nu+n}\ |\ n\in\mathbb{Z}_{\geq0}\right\}  $ and $\left\{  e^{-i\nu
x}J_{\nu-n}\ |\ n\in\mathbb{Z}_{\geq0}\right\}  $, respectively.\footnote{Yes,
$n=0$ is in both sectors, as defined. \ If this is troubling, then feel free
to define \emph{three} sectors to avoid any overlap.} \ When $\nu
\notin\mathbb{Z}$, $J_{\nu\pm n}$ give two independent states for all $n>0$.
\ Thus \emph{the number of states has doubled} from the $\nu=0$ situation, no
matter how small $\nu\neq0$ is.\footnote{This $2\rightarrow1$ accounting of
the states, as $\nu\rightarrow0$, is complementary to the $1\rightarrow2$
state counting for linear \textquotedblleft wall\textquotedblright%
\ potentials, as the slope of the walls goes to zero.} \ The effect of
$\nu\notin\mathbb{Z}$ is to restore \emph{all} left-moving plane waves to the
span of the energy eigenfunctions, in sharp contrast to $\nu=0$ where only the
right-moving plane waves could be expanded in $J_{n}\left(  me^{ix}\right)  $. \ 

So the energies are given on the two sectors by%
\begin{equation}
E_{\pm n}=\left(  \nu\pm n\right)  ^{2}\text{ \ \ for \ \ }n=0,1,\cdots
\end{equation}
For generic $\nu$, $E_{\pm n}$\ are not degenerate. \ But if $\nu
-1/2=\lambda\in\mathbb{Z}$, eigenvalues are doubly degenerate. \ For example,
if $\nu=1/2$, the degenerate pairs of energies are $E_{0,-1}$, $E_{1,-2}$,
etc. \ In this case there are two independent states for \emph{each} energy.

Otherwise, when $\nu\in\mathbb{Z}$, most if not all of the $J_{\nu-n}$\ states
\emph{meld}\ into the $J_{\nu+n}$\ states\footnote{That is to say, integer
$\nu$\ are \emph{exceptional} spectral points. \ See \cite{Kato,Heiss}.},
since the Bessels with negative integer index are not independent of those
with positive integer index, i.e. $J_{\nu-n}\left(  z\right)  =\left(
-1\right)  ^{\nu-n}J_{n-\nu}\left(  z\right)  $, for $n$ and $\nu\in
\mathbb{Z}$. \ Sorting out the independent functions leaves just $\left\{
J_{\lambda}\ |\ \lambda\in\mathbb{Z}_{\geq-\nu}\right\}  $. \ For instance,
when $\nu\in\mathbb{Z}_{\geq0}$\ the independent states are $e^{-i\nu x}%
J_{\nu-n}$ for $\nu-n\in\left\{  0,1,\cdots,\nu\right\}  $ (alternatively
$n\in\left\{  \nu,\nu-1,\cdots,1,0\right\}  $) and $e^{-i\nu x}J_{\nu+n}$ for
$n\in\mathbb{Z}_{\geq1}$. \ That is to say, when $\nu\in\mathbb{Z}_{\geq0}$,
the independent states are nondegenerate with energies $0,1,4,9,\cdots$,
exactly as in the $\nu=0$ situation\ (although the wave functions are
different for $\nu=0$ and $\nu\neq0$). \ Moreover, when $\nu\in\mathbb{Z}%
_{\geq0}$\ only a finite number of left-moving plane waves are in the span of
the energy eigenfunctions, namely: $\ \left\{  e^{-i\nu x},e^{-i\left(
\nu-1\right)  x},\cdots,e^{-ix}\right\}  $. Similar results apply when $\nu
\in\mathbb{Z}_{<0}$, although in this case only right-moving plane waves are
spanned by the eigenfunctions, namely: \ $\left\{  e^{i\left\vert
\nu\right\vert x},e^{i\left(  1+\left\vert \nu\right\vert \right)
x},e^{i\left(  2+\left\vert \nu\right\vert \right)  x}\cdots\right\}  $, with
the first $\left\vert \nu\right\vert $ right-movers absent from the span.
\ But again, the energies are $0,1,4,9,\cdots$, exactly as in the $\nu=0$
situation. \ In any case, all of these situations involving integer $\nu$ are
(gauge) equivalent through energy preserving one-to-one maps connecting the
various Bessel functions. \ Map details are left to the reader to establish.

For generic $\nu$, the dual space may be constructed for all eigenfunctions,
but again \emph{not} by the naive PT method. \ Every function in the right
sector, $e^{-i\nu x}J_{\nu+n}\left(  me^{ix}\right)  $, is once again
self-orthogonal (except $n=0$) as is every function in the left sector. \ So,
an appropriate dual function is not proportional to the eigenfunction, rather
it is given by $e^{i\nu x}J_{-\nu-n}\left(  me^{ix}\right)  /C_{n}$ where
$C_{n}=\frac{1}{2\pi}\int_{0}^{2\pi}J_{-\nu-n}\left(  me^{ix}\right)
J_{\nu+n}\left(  me^{ix}\right)  dx$. \ This dual function is not an
eigenfunction of $H$. \ Instead, it is an eigenfunction of
\begin{equation}
\widetilde{H}=\left(  -i\partial_{x}-\nu\right)  ^{2}+m^{2}e^{2ix}%
\end{equation}
obtained by flipping the sign of the vector potential\footnote{Note this is
not what would happen under CPT, where \textquotedblleft C\textquotedblright%
\ here \emph{really is} charge conjugation, since $\nu$ $\propto eA$ and this
does \emph{not} change sign under CPT.}, with the same eigenvalue as the
associated right-sector function,
\begin{equation}
\widetilde{H}\ e^{i\nu x}J_{-\nu-n}\left(  me^{ix}\right)  =\left(
\nu+n\right)  ^{2}e^{i\nu x}J_{-\nu-n}\left(  me^{ix}\right)
\end{equation}
Similarly the dual for every left-sector eigenfunction, say $e^{-i\nu x}%
J_{\nu-n}\left(  me^{ix}\right)  $, is $e^{i\nu x}J_{-\nu+n}\left(
me^{ix}\right)  /D_{n}$ where $D_{n}=\frac{1}{2\pi}\int_{0}^{2\pi}J_{-\nu
+n}\left(  me^{ix}\right)  J_{\nu-n}\left(  me^{ix}\right)  dx$, and
$\widetilde{H}\ e^{i\nu x}J_{-\nu+n}\left(  me^{ix}\right)  =\left(
\nu-n\right)  ^{2}e^{i\nu x}J_{-\nu+n}\left(  me^{ix}\right)  $. \ The duals
for right- and left-sector eigenfunctions are guaranteed to be orthogonal to
left- and right-sector eigenfunctions, respectively, for generic $\nu$ where
$E_{\pm n} $\ are not degenerate, by standard arguments (as in
(\ref{HActingOnDuals}) et seq.). \ But if $\nu-1/2=\lambda\in\mathbb{Z}$,
further consideration is needed, since right- and left-sector eigenvalues are
degenerate. \ An adaptation to biorthogonal systems of the usual degenerate
state methods in conventional QM suffices to handle this situation for both
right and left sectors, in a straightforward way. \ 

Canonical integral representations for the wave functions and their duals may
also be exhibited for this model, as given for example by (\ref{Sonine}) with
$\sqrt{E}$ replaced by $\nu\pm n$ and $-\nu\pm n.$

Rather than dwell on this, for simplicity we now limit our discussion to the
right sector for any $\nu\notin\mathbb{Z}_{<0}$. \ This is an invariant
subspace spanned by $\left\{  e^{-i\nu x}J_{\nu+n}\left(  me^{ix}\right)
\ |\ n=0,1,\cdots\right\}  $, and as previously pointed out, all right-moving
plane waves\ are contained in this subspace for $\nu\notin\mathbb{Z}_{<0}$.
\ As an alternative to calling it the right sector, it is the \emph{analytic
sector} in the variable $z=me^{ix}$. \ For this subspace, rather than as
eigenfunctions of $\widetilde{H}$, the duals may be constructed as finite
polynomials in $1/z$ in a manner that closely parallels our previous analysis
of the $\nu=0$ situation for $2\pi$-periodic functions.

\paragraph{Gegenbauer's generalization of the Neumann polynomials and the
corresponding construction of the dual space}

In this generalization, the index $n$ is always $\geq0$ and integral, while
the index $\nu$ is allowed any value \emph{except} $-1,-2,-3,\cdots$.
\ Negative $\nu$ are allowed, just so long as they are \emph{not} integral.
\ An explicit definition is (\cite{Watson} \S 9.2)%
\begin{align}
A_{n,\nu}\left(  w\right)   &  =\frac{2^{\nu+n}\left(  \nu+n\right)  }%
{w^{n+1}}\sum_{k=0}^{\left\lfloor n/2\right\rfloor }\frac{\Gamma\left(
\nu+n-k\right)  }{k!}\left(  \frac{w}{2}\right)  ^{2k}\\
&  =\frac{2^{\nu+n}\Gamma\left(  \nu+n+1\right)  }{w^{n+1}}\left(
1+\frac{w^{2}/4}{\left(  \nu+n-1\right)  }+\frac{w^{4}/16}{2!\left(
\nu+n-1\right)  \left(  \nu+n-2\right)  }+\cdots\right) \nonumber
\end{align}
The Neumann polynomials are a special case: $\ A_{n,\nu=0}\left(  w\right)
=\varepsilon_{n}O_{n}\left(  w\right)  $. \ (NB $A_{n,\nu=0}\left(  w\right)
=\frac{1}{w}A_{n}\left(  w\right)  $ in \emph{our} previous notation. \ Sorry
about that!) \ The Fourier transform on a simple counter-clockwise contour
enclosing the origin once is
\begin{equation}
\frac{1}{2\pi i}%
%TCIMACRO{\toint }%
%BeginExpansion
{\textstyle\oint}
%EndExpansion
e^{izw}A_{n,\nu}\left(  w\right)  dw=2^{\nu}i^{n}\left(  \nu+n\right)
\Gamma\left(  \nu\right)  C_{n}^{\nu}\left(  z\right)  \text{ \ \ where
\ \ }\frac{1}{\left(  1-2tz+t^{2}\right)  ^{\nu}}\equiv\sum_{n=0}^{\infty
}t^{n}C_{n}^{\nu}\left(  z\right)
\end{equation}
That is to say, the contour Fourier transforms of Gegenbauer's polynomials
$A_{n,\nu}$ are just \emph{the} Gegenbauer polynomials $C_{n}^{\nu}$.

The elementary Cauchy kernel is now given by%
\begin{equation}
\frac{1}{w-z}=\sum_{n=0}^{\infty}A_{n,\nu}\left(  w\right)  \ z^{-\nu}%
J_{\nu+n}\left(  z\right)  \ \ \text{for}\ \ \left\vert z\right\vert
<\left\vert w\right\vert
\end{equation}
This follows from expanding $\frac{z^{\nu}}{w-z}$ in powers of $z$ and writing
$z^{\mu}$ for non-negative-integer $\mu$ in terms of Bessels\ (\cite{Watson}
\S 5.2).
\begin{equation}
z^{\mu}=\sum_{n=0}^{\infty}\frac{\left(  \mu+2n\right)  2^{\mu}\Gamma\left(
\mu+n\right)  }{n!}J_{\mu+2n}\left(  z\right)  \text{ \ \ for \ \ }\mu
\notin\mathbb{Z}_{<0}%
\end{equation}
Other kernels may also be constructed in closed form (for example, see
(\ref{nuJKernel}) and (\ref{nuHKernel}) below).

These polynomials may be used to construct the dual space for the right
sector. \ To this end we define for all $n\in\mathbb{Z}_{\geq0}$
\begin{align}
\psi_{n,\nu}\left(  me^{ix}\right)   &  \equiv Z_{n}\frac{e^{-i\nu x}}{m^{\nu
}}J_{\nu+n}\left(  me^{ix}\right) \nonumber\\
\chi_{n,\nu}\left(  me^{ix}\right)   &  \equiv\frac{me^{ix}}{Z_{n}}A_{n,\nu
}\left(  me^{ix}\right)
\end{align}
where $Z_{n}$ is a normalization to be chosen for convenience, later (see
(\ref{NormChoice})). \ As explicit series%
\begin{align}
\psi_{n,\nu}\left(  me^{ix}\right)   &  =\frac{Z_{n}}{2^{\nu+n}}m^{n}%
e^{inx}\sum_{k=0}^{\infty}\frac{\left(  -1\right)  ^{k}}{k!\Gamma\left(
1+k+\nu+n\right)  }\left(  \frac{m}{2}\right)  ^{2k}e^{2ikx}\nonumber\\
\chi_{n,\nu}\left(  me^{ix}\right)   &  =\frac{2^{\nu+n}\left(  \nu+n\right)
}{Z_{n}m^{n}}e^{-inx}\sum_{k=0}^{\left\lfloor n/2\right\rfloor }\frac
{\Gamma\left(  \nu+n-k\right)  }{k!}\left(  \frac{m}{2}\right)  ^{2k}e^{2ikx}
\label{nuSeries}%
\end{align}
Both sets of functions are clearly $2\pi$-periodic. \ The energy
eigenfunctions obey Schr\"{o}dinger's equation%
\begin{equation}
H\psi_{n,\nu}\left(  me^{ix}\right)  =\left(  -i\frac{d}{dx}+\nu\right)
^{2}\psi_{n,\nu}\left(  me^{ix}\right)  +m^{2}e^{2ix}\psi_{n,\nu}\left(
me^{ix}\right)  =\left(  \nu+n\right)  ^{2}\psi_{n,\nu}\left(  me^{ix}\right)
\end{equation}
while the finite polynomials obey the inhomogeneous differential equation%
\begin{align}
\left(  \widetilde{H}-\left(  \nu+n\right)  ^{2}\right)  \chi_{n,\nu}\left(
me^{ix}\right)   &  =\left(  -i\frac{d}{dx}-\nu\right)  ^{2}\chi_{n,\nu
}\left(  me^{ix}\right)  +\left\{  m^{2}e^{2ix}-\left(  \nu+n\right)
^{2}\right\}  \chi_{n,\nu}\left(  me^{ix}\right) \nonumber\label{HnuOnDuals}\\
&  =\frac{2^{\nu}\left(  \nu+n\right)  }{Z_{n}}\left\{
\begin{array}
[c]{c}%
\frac{\Gamma\left(  \nu+\frac{1}{2}\left(  n+1\right)  \right)  }%
{\Gamma\left(  1+\frac{1}{2}\left(  n-1\right)  \right)  }\times2me^{ix}\text{
\ for }n\text{ odd}\\
\\
\frac{\Gamma\left(  \nu+\frac{1}{2}n\right)  }{\Gamma\left(  1+\frac{1}%
{2}n\right)  }\times m^{2}e^{2ix}\text{ \ for\ }n\text{ even}%
\end{array}
\right.
\end{align}
The two sets of functions satisfy the biorthonormality condition%
\begin{equation}
\frac{1}{2\pi}\int_{0}^{2\pi}\chi_{k,\nu}\left(  me^{ix}\right)  \psi_{n,\nu
}\left(  me^{ix}\right)  dx=\delta_{kn}\text{ \ \ for \ \ }k,n\in
\mathbb{Z}_{\geq0}%
\end{equation}
The proof for $k=n$ is obvious from the series (\ref{nuSeries}). \ The proof
for $k\neq n$ proceeds by first acting with $H$ on $\psi_{n,\nu}$ under the
integral, then integrating by parts to act with $\widetilde{H}$ on
$\chi_{k,\nu}$, and finally comparing the results, i.e. just following the
same line of argument as in the case of a single exponential potential,
(\ref{HActingOnDuals}) et seq. \ But note the importance of the $\nu$ sign
change for $\widetilde{H}$ compared to $H$.

Thus for any $\nu\notin\mathbb{Z}_{<0}$ we have obtained for the right sector
a biorthogonal system that is manifestly in one-to-one correspondence to that
of the $\nu=0$ situation: \ $\left\{  \chi_{k,\nu},\psi_{n,\nu}\right\}
\longleftrightarrow\left\{  A_{k},J_{n}\right\}  $. \ We leave the
construction of explicit integral transformations between the $\nu\neq0$ and
the $\nu=0$ systems as an exercise. \ (Hint: \ see \cite{Watson} \S 11.3)

On the dual space of the right sector, the bilocal norm kernel can once again
be obtained in closed form if we choose
\begin{equation}
Z_{n}^{2}=\frac{\sqrt{4\pi}}{2^{\nu}\Gamma\left(  \nu+1/2\right)  }%
\frac{\left(  \nu+n\right)  \Gamma\left(  2\nu+n\right)  }{n!}
\label{NormChoice}%
\end{equation}
Then the hermitian kernel is%
\begin{equation}
J\left(  x,y\right)  =\sum_{n=0}^{\infty}\psi_{n,\nu}\left(  me^{-ix}\right)
\psi_{n,\nu}\left(  me^{iy}\right)  =\frac{J_{\nu}\left(  me^{-ix}%
-me^{iy}\right)  }{\left(  me^{-ix}-me^{iy}\right)  ^{\nu}}%
\end{equation}
This follows from the addition theorem (\cite{Watson} \S 11.4)%
\begin{equation}
\frac{J_{\nu}\left(  w-z\right)  }{\left(  w-z\right)  ^{\nu}}=\frac
{\sqrt{4\pi}}{2^{\nu}\Gamma\left(  \nu+1/2\right)  }\sum_{n=0}^{\infty}%
\frac{\left(  \nu+n\right)  \Gamma\left(  2\nu+n\right)  }{n!}w^{-\nu}%
J_{\nu+n}\left(  w\right)  z^{-\nu}J_{\nu+n}\left(  z\right)
\label{nuJKernel}%
\end{equation}
Note that the ratio $\frac{J_{\nu}\left(  w-z\right)  }{\left(  w-z\right)
^{\nu}}$ is analytic in $w$ and $z$ for all $\nu$. \ $J\left(  x,y\right)  $
clearly reduces to the previous case (\ref{JKernel})\ when $\nu=0$.

It is straightforward to act on this norm kernel with the Hamiltonian to
obtain a closed form for the Hamiltonian kernel on the dual space, as we did
earlier for $\nu=0$. \ We find%
\begin{align}
H\left(  x,y\right)   &  \equiv\left(  \left(  -i\partial_{y}+\nu\right)
^{2}+m^{2}e^{2iy}\right)  \frac{J_{\nu}\left(  me^{-ix}-me^{iy}\right)
}{\left(  me^{-ix}-me^{iy}\right)  ^{\nu}}\nonumber\\
&  =\nu^{2}J\left(  x,y\right)  +\left(  1+2\nu\right)  m^{2}e^{-ix+iy}%
\ \frac{J_{\nu+1}\left(  me^{-ix}-me^{iy}\right)  }{\left(  me^{-ix}%
-me^{iy}\right)  ^{\nu+1}}\nonumber\\
&  =\sum_{n=0}^{\infty}\left(  \nu+n\right)  ^{2}\psi_{n,\nu}\left(
me^{-ix}\right)  \psi_{n,\nu}\left(  me^{iy}\right)  \label{nuHKernel}%
\end{align}
Once again, this is an hermitian kernel,\ $H\left(  x,y\right)  =H^{\ast
}\left(  y,x\right)  $, and it smoothly reduces to the flux-free $\Phi=0$ case
when $\nu=0$, as in (\ref{DualHamiltonianKernel}).

These kernels may be used to compute norms and energy averages, just as in
(\ref{JNorm}) and (\ref{HAverage}), for arbitrary states in the right sector,%
\begin{equation}
\psi\left(  x\right)  =\sum_{n=0}^{\infty}c_{n}\psi_{n,\nu}\left(
me^{ix}\right)
\end{equation}
through the use of the corresponding dual functions
\begin{equation}
\psi_{\text{dual}}\left(  x\right)  =\sum_{n=0}^{\infty}c_{n}^{\ast}%
\chi_{n,\nu}\left(  me^{ix}\right)
\end{equation}
Other kernels and corresponding expectation values can be easily constructed,
including some like $-i\partial_{y}J\left(  x,y\right)  $ which are
non-hermitian. \ But we will not pursue this here.

\section{Two exponentials}

This is a complex version of the Morse potential \cite{MorsePotl}. \ With a
constant vector potential included, the Hamiltonian is
\begin{equation}
H=\left(  p+\nu\right)  ^{2}+\mu_{1}\exp\left(  ix\right)  +\mu_{2}\exp\left(
2ix\right)
\end{equation}
The analysis of the classical solutions proceeds as before, for fixed energy,
and yields trajectories in terms of known functions. \ We forego further
discussion of the classical motion (except in a special case given below) and
consider immediately the quantum system.

The corresponding Schr\"{o}dinger energy eigenvalue equation is%
\begin{equation}
\left[  \left(  -i\frac{d}{dx}+\nu\right)  ^{2}+\mu_{1}\exp\left(  ix\right)
+\mu_{2}\exp\left(  2ix\right)  \right]  \psi_{E}\left(  x\right)  =E\psi
_{E}\left(  x\right)
\end{equation}
Single-valued solutions, bounded on the real line, exist for any $E\geq0$ and
are given by%
\begin{align}
\psi_{E}^{\pm}\left(  x\right)   &  =z^{-\nu-\frac{1}{2}}%
\operatorname{WhittakerM}\left(  \frac{1}{2}\frac{\mu_{1}}{\sqrt{-\mu_{2}}%
},\pm\sqrt{E},2\sqrt{-\mu_{2}}z\right)  \ \ \ \ \ \text{where \ \ \ \ }z\equiv
e^{ix}\nonumber\\
&  =e^{\left(  -\nu\pm\sqrt{E}\right)  ix}e^{-\sqrt{-\mu_{2}}e^{ix}}M\left(
\frac{1}{2}\pm\sqrt{E}-\frac{1}{2}\frac{\mu_{1}}{\sqrt{-\mu_{2}}},1\pm
2\sqrt{E},2\sqrt{-\mu_{2}}e^{ix}\right)  \label{Psi2Exp}%
\end{align}
These eigenfunctions are all manifestly even under PT, for $\nu$, $\mu_{1}$,
and $\sqrt{-\mu_{2}}$ all real. \ As indicated by the notation, there is a
double degeneracy for generic $E$. \ Here $M\left(  a,b,z\right)
\equiv\operatorname{KummerM}\left(  a,b,z\right)  $ is Kummer's confluent
hypergeometric function (\cite{Abram} Chapter 13) also known as $\left.
_{1}F_{1}\right.  \left(  a;b;z\right)  $. \
\begin{equation}
M\left(  a,b,z\right)  =\left.  _{1}F_{1}\right.  \left(  a;b;z\right)
=\sum_{n=0}^{\infty}\frac{\Gamma\left(  a+n\right)  }{\Gamma\left(  a\right)
}\frac{\Gamma\left(  b\right)  }{\Gamma\left(  b+n\right)  }\frac{z^{n}}{n!}%
\end{equation}
So the quantum energy spectrum of this Hamiltonian is the positive real line
for any choice of $\mu_{1}$\ and $\mu_{2}$. \ Since it is not always possible
to shift $x$ to obtain both real $\mu_{1}$ and real $\mu_{2}$, simultaneously,
this model is not translationally equivalent to a PT symmetric theory, in
general, and therefore the model provides an interesting counter-example to
the naive dictate that real energies require PT symmetry.\footnote{However,
the model is clearly continuously connected to PT invariant theories through
\emph{independent} variations of both $\mu_{1}$ and $\mu_{2} $. \ All such
variations are \emph{isospectral deformations} of $H$, i.e. they do not change
the total set of energy eigenvalues even though they change the form of $H$
and individual energy eigenfunctions. \ Relatedly, such variations define
continuous \emph{deformations} of $\mathcal{PT}$. \ This statement should be
compared to various theorems in the literature on the interplay between
general anti-linear symmetries and the reality of the energy spectrum
\cite{BenderParity,MostaJMP2,MostaJMP4,Blasi}.} \ 

The model defines a biorthogonal system for all $E\geq0$, and for generic $E $
the dual space can be constructed using the methods of the previous example
involving only one exponential. \ In particular, if we restrict our attention
to eigenfunctions periodic on $x\in\left[  0,2\pi\right]  $, we again find
exceptional biorthogonal systems. \ Due to the pre-factor, $e^{\left(  -\nu
\pm\sqrt{E}\right)  ix}$, $2\pi$-periodicity requires $-\nu\pm\sqrt{E}%
=n\in\mathbb{Z}$. \ So the energies are given in pairs by%
\begin{equation}
E_{\pm n}=\left(  \nu\pm n\right)  ^{2}\text{ \ \ for \ \ }n=0,1,\cdots
\end{equation}
This is exactly the same as in the previous model involving only one
exponential. \ For generic $\nu$ the eigenfunctions again define left and
right sectors, as in the single exponential case. \ The dual space can be
constructed accordingly.

If we again restrict our attention to the right sector, for simplicity, the
dual space may be constructed from polynomials in $1/z$. \ For general
$\mu_{1}$ and $\mu_{2}$ the relevant polynomials have been investigated by
Erd\'{e}lyi \cite{Erdelyi}. \ We defer a discussion of their structure for
general $\mu_{1}$ and $\mu_{2}$ until the next section of the paper, where we
give a unified treatment for all Hamiltonians of the form (\ref{H}). \ For
special choices of $\mu_{1}\neq0\neq\mu_{2}$, however, the biorthogonal
structure simplifies dramatically, and many results can again be obtained in
closed form. \ If we set $\nu=0$ and if we make a particular choice for the
coefficients of the two exponentials, somewhat surprisingly the eigenfunctions
can again be expressed in terms of Bessel functions.

\paragraph{Simplification for a particular two-exponential case}

Let%
\begin{equation}
H=p^{2}+me^{ix}-m^{2}e^{2ix} \label{H2ExpSimp}%
\end{equation}
Analysis of the classical solutions yields trajectories in terms of elementary
functions as in the case of one exponential. \ For $E>0$, the complex momentum
trajectories are now \textquotedblleft lima\c{c}ons\textquotedblright\ (for
examples, see \cite{MathWorld}) centered on the $\operatorname{Re}p$ axis,
rather than the circles of the previous case involving only one exponential.
\ There is also considerably more structure in the $E=0 $ classical solutions.
Simple special cases occur when $e^{-ix\left(  0\right)  }=m$, for which cases
we encounter turning points. \ 

We now consider the quantum system and immediately note the Darboux
factorization \cite{Darboux} of the Hamiltonian.%
\begin{equation}
H=-\frac{d^{2}}{dx^{2}}+me^{ix}-m^{2}e^{2ix}=\left(  -i\frac{d}{dx}%
-me^{ix}\right)  \left(  -i\frac{d}{dx}+me^{ix}\right)  \label{Hfactored}%
\end{equation}
Schr\"{o}dinger's eigenvalue equation, the allowed energy eigenvalues, and the
periodic eigenfunctions for $x\in\left[  0,2\pi\right]  $ are now given by
\begin{equation}
-\frac{d^{2}}{dx^{2}}\psi_{n}\left(  x\right)  +\left(  me^{ix}-m^{2}%
e^{2ix}\right)  \psi_{n}\left(  x\right)  =n^{2}\psi_{n}\left(  x\right)
\end{equation}%
\begin{align}
\psi_{n}\left(  x\right)   &  =Z_{n}\left(  \frac{me^{ix}}{2}\right)
^{n}\left\{  \sum_{k=0}^{\infty}\frac{1}{k!\Gamma\left(  k+\frac{1}%
{2}+n\right)  }\left(  \frac{me^{ix}}{2}\right)  ^{2k}-\sum_{k=0}^{\infty
}\frac{1}{k!\Gamma\left(  k+\frac{3}{2}+n\right)  }\left(  \frac{me^{ix}}%
{2}\right)  ^{2k+1}\right\} \nonumber\\
&  =Z_{n}\sqrt{\frac{me^{ix}}{2}}\left\{  I_{n-1/2}\left(  me^{ix}\right)
-I_{n+1/2}\left(  me^{ix}\right)  \right\} \nonumber\\
&  =\frac{Z_{n}}{\sqrt{2}}\left(  -i\frac{d}{dx}-me^{ix}+1+n\right)
\frac{I_{n+1/2}\left(  me^{ix}\right)  }{\sqrt{me^{ix}}}%
\end{align}
for all integer $n$, both positive and negative. \ $Z_{n}$ is a normalization
constant.\footnote{The choice $Z_{n}=4^{n}\Gamma\left(  \frac{1}{2}+n\right)
$ gives a balanced asymptotic behavior for both bilocal norm kernels,
$J\left(  x,y\right)  $ and $K\left(  x,y\right)  $. \ } \ The $I$s are
modified Bessel functions, as given by the explicit series (\ref{Imu}), or,
because in this case they have semi-integer indices, as given by finite sums
of powers multiplying hyperbolic functions of $me^{ix}$ (\cite{Abram}
\S 10.2). \ For $n\in\mathbb{Z}_{n\geq0}$
\begin{equation}
I_{n+1/2}\left(  z\right)  =\sqrt{\frac{2}{\pi}}z^{n+1/2}\left(  \frac{1}%
{z}\frac{d}{dz}\right)  ^{n}\frac{\sinh z}{z}\ ,\ \ \ I_{-n-1/2}\left(
z\right)  =\sqrt{\frac{2}{\pi}}z^{n+1/2}\left(  \frac{1}{z}\frac{d}%
{dz}\right)  ^{n}\frac{\cosh z}{z}%
\end{equation}
The non-degenerate ground state is simply $\psi_{n=0}\left(  x\right)
=Z_{0}e^{-me^{ix}}/\sqrt{\pi}$. \ The result $H\psi_{n=0}\left(  x\right)  =0
$ follows immediately from $\left(  -i\frac{d}{dx}+me^{ix}\right)
e^{-me^{ix}}=0 $ and the factorized form (\ref{Hfactored}), just as in
supersymmetric QM \cite{Darboux}.

Duals for all the energy eigenfunctions may be constructed from linear
combinations of $\psi_{n\leq0}$ and $\psi_{n\geq0}$, similar to
(\ref{Eigenfunctions}). \ On the other hand, if we restrict to the right
sector, i.e. the subspace spanned by $\left\{  \psi_{n}\ |\ n\geq0\right\}  $,
we may once again make use of only finite polynomials in $e^{-ix}$ to
construct the dual space. \ Explicit calculation gives%
\begin{align}
\chi_{n}\left(  x\right)   &  =\frac{1}{Z_{n}}\left(  \frac{2}{me^{ix}%
}\right)  ^{n}\left\{  \sum_{k=0}^{\left\lfloor n/2\right\rfloor }%
\frac{\left(  -1\right)  ^{k}\Gamma\left(  n-k+\frac{1}{2}\right)  }%
{k!}\left(  \frac{me^{ix}}{2}\right)  ^{2k}+\sum_{k=0}^{\left\lfloor \left(
n-1\right)  /2\right\rfloor }\frac{\left(  -1\right)  ^{k}\Gamma\left(
n-k-\frac{1}{2}\right)  }{k!}\left(  \frac{me^{ix}}{2}\right)  ^{2k+1}\right\}
\nonumber\\
&  =\frac{1}{Z_{n}}\frac{ime^{ix}}{\sqrt{2}}\left\{  \frac{i^{n}}{\left(
n+1/2\right)  }A_{n,1/2}\left(  ime^{ix}\right)  +\frac{i^{n-1}}{\left(
n-1/2\right)  }A_{n-1,1/2}\left(  ime^{ix}\right)  \right\}
\end{align}
These $\left\{  \chi_{n}\ |\ n\geq0\right\}  $ are biorthonormalized duals for
$\left\{  \psi_{n}\ |\ n\geq0\right\}  $. \ By construction, we have%
\begin{equation}
\frac{1}{2\pi}\int_{0}^{2\pi}\chi_{k}\left(  x\right)  \psi_{j}\left(
x\right)  =\delta_{kj}%
\end{equation}
for all $j,k\geq0$. \ The dual functions are now solutions to the
inhomogeneous equations%
\begin{equation}
-\frac{d^{2}}{dx^{2}}\chi_{n}\left(  x\right)  +\left(  me^{ix}-m^{2}%
e^{2ix}\right)  \chi_{n}\left(  x\right)  -n^{2}\chi_{n}\left(  x\right)
=\alpha_{n}m^{2}e^{2ix}+\beta_{n}me^{ix}%
\end{equation}
where the coefficients of the inhomogeneous terms are given by%
\begin{align}
\alpha_{n=2k}  &  =\frac{\left(  -1\right)  ^{k+1}\Gamma\left(  k+\frac{1}%
{2}\right)  }{k!Z_{2k}}\ ,\ \ \ \beta_{n=2k}=-\left(  2k+1\right)
\alpha_{n=2k}\nonumber\\
\alpha_{n=2k+1}  &  =\frac{\left(  -1\right)  ^{k+1}\Gamma\left(  k+\frac
{1}{2}\right)  }{k!Z_{2k+1}}\ ,\ \ \ \beta_{n=2k+1}=2k\alpha_{n=2k+1}%
\end{align}
The result $\int_{0}^{2\pi}\chi_{k}\left(  x\right)  \psi_{j}\left(  x\right)
dx=0$ for $k\neq j$ again follows from integrating by parts and from $\int
_{0}^{2\pi}e^{ix}\psi_{j}\left(  x\right)  dx=0=\int_{0}^{2\pi}e^{2ix}\psi
_{j}\left(  x\right)  dx$ for all $j\geq0$, as in (\ref{HActingOnDuals}) et
seq. \ 

The construction of the bilocal norm and Hamiltonian kernels proceeds as
before. \ Acting on the dual space of the right sector, we find ($w\equiv
me^{-ix},\ z\equiv me^{iy}$)%
\begin{align}
J\left(  x,y\right)   &  =\sum_{n=0}^{\infty}\overline{\psi_{n}\left(
x\right)  }\psi_{n}\left(  y\right) \label{J2Exp}\\
&  =\frac{1}{2\sqrt{wz}}\left[  \left(  w\frac{d}{dw}-w\right)  \left(
z\frac{d}{dz}-z\right)  +\frac{1}{4}\right]  \sum_{n=0}^{\infty}\left\vert
Z_{n}\right\vert ^{2}I_{n+1/2}\left(  w\right)  I_{n+1/2}\left(  z\right)
\nonumber\\
&  +\frac{1}{2\sqrt{wz}}\sum_{n=0}^{\infty}\left\vert Z_{n}\right\vert
^{2}n\left(  n+1\right)  I_{n+1/2}\left(  w\right)  I_{n+1/2}\left(  z\right)
\nonumber\\
&  +\frac{1}{2\sqrt{wz}}\left[  w\frac{d}{dw}-w+z\frac{d}{dz}-z\right]
\sum_{n=0}^{\infty}\left\vert Z_{n}\right\vert ^{2}\left(  n+\frac{1}%
{2}\right)  I_{n+1/2}\left(  w\right)  I_{n+1/2}\left(  z\right) \nonumber
\end{align}
The choice $\left\vert Z_{n}\right\vert ^{2}=\left(  2n+1\right)  \pi$
facilitates evaluation of the sums, since (\cite{Watson} \S 11.41 Eqn(6))
\begin{equation}
\frac{\sinh\sqrt{w^{2}+z^{2}-2wzs}}{\sqrt{w^{2}+z^{2}-2wzs}}=\sqrt{\frac{\pi
}{2}}\frac{I_{1/2}\left(  \sqrt{w^{2}+z^{2}-2wzs}\right)  }{\sqrt[4]%
{w^{2}+z^{2}-2wzs}}=\frac{\pi}{\sqrt{wz}}\sum_{n=0}^{\infty}\left(
n+1/2\right)  I_{n+1/2}\left(  w\right)  I_{n+1/2}\left(  z\right)  \left(
-1\right)  ^{n}P_{n}\left(  s\right)
\end{equation}
This result, along with $\left(  -1\right)  ^{n}\left.  P_{n}\left(  s\right)
\right\vert _{s=-1}=1$ and $2\left(  -1\right)  ^{n}\frac{d}{ds}\left.
P_{n}\left(  s\right)  \right\vert _{s=-1}=-n\left(  n+1\right)  $, leads to%
\begin{align}
\frac{\pi}{\sqrt{wz}}\sum_{n=0}^{\infty}\left(  n+1/2\right)  I_{n+1/2}\left(
w\right)  I_{n+1/2}\left(  z\right)   &  =\frac{\sinh\left(  w+z\right)
}{w+z}\\
\frac{\pi}{\sqrt{wz}}\sum_{n=0}^{\infty}\left(  n+1/2\right)  n\left(
n+1\right)  I_{n+1/2}\left(  w\right)  I_{n+1/2}\left(  z\right)   &  =\left(
\cosh\left(  z+w\right)  -\frac{\sinh\left(  z+w\right)  }{\left(  z+w\right)
}\right)  \frac{2wz}{\left(  w+z\right)  ^{2}}\nonumber
\end{align}
Unfortunately, for the time being, a utilitarian closed form for the last sum
in (\ref{J2Exp}) eludes us. \ We have only found an integral representation
for it (see \cite{Prud}, Vol. II, \textbf{5.7.17,} Eqn(11), p 675). \ 

Given a sufficiently simple result for that third sum, a closed form for the
bilocal Hamiltonian $H\left(  x,y\right)  $ would then follow by acting with
$H $ on $J\left(  x,y\right)  $, as in the previous cases involving one exponential.

Canonical integral representations for the wave functions, and their duals,
can also be exhibited for the two exponential model, especially in the
simplified case (\ref{H2ExpSimp}). \ The easiest of these is probably
Poisson's integral (see \cite{Abram}\ \textbf{10.1.13} and \textbf{10.1.14}
with\textbf{\ }$z\rightarrow iz$), or alternatively just (\ref{Sonine}) with
$\sqrt{E}$ replaced by $n+1/2$ and $m\rightarrow im$ in the argument of the
Bessel function. \ Corresponding integral representations for the duals follow
from biorthonormality, for given choices of the kernels (e.g. for Poisson's
integral, an appropriate kernel is given by \cite{Abram}\ \textbf{10.2.36}).
\ This allows construction of a Table as in the case of the one exponential
model. \ We leave the details of this construction to the well-motivated,
interested reader.

\section{Biorthogonal systems for $H=p^{2}+\sum_{k>0}\mu_{k}\exp\left(
ikx\right)  $}

We indicate methods to analyze the general case, reserving some of the details
to be presented elsewhere. \ For $x\in\left(  -\infty,+\infty\right)  $
without any periodicity constraints, the spectrum of this Hamiltonian is known
\cite{Gasymov} to be the real, positive half-line, for \emph{any} choice of
$\mu$s such that $\sum_{k>0}\left\vert \mu_{k}\right\vert <\infty$, and not
just for those $\mu$s which are real or $x$-translationally equivalent to real
values. \ So PT symmetry is not required for real energy eigenvalues in this
model. \ Here, we will restrict our attention to $2\pi$-periodic functions and
their duals, to obtain a discrete subset of real energy eigenvalues, namely
just $\left\{  E_{n}=n^{2}\ |\ n=0,1,\cdots\right\}  $. \ 

A simple procedure to construct this biorthogonal system is to \emph{begin}
with the dual functions.\footnote{So these systems are an exception to the
statement: \ \textquotedblleft A direct and frontal attack on the problem of
determining [the dual polynomials] is not usually fruitful.\textquotedblright%
\ -- Morse and Feshbach \cite{Morse} p 931.} \ We assume these to be
polynomials in powers of $z^{-1}\equiv e^{-ix}$, so we write
\begin{equation}
\chi_{n}\left(  z\right)  =\frac{1}{z^{n}}\sum_{j=0}^{n}c_{n,j}z^{j}%
\ ,\ \ \ H=\left(  z\frac{d}{dz}\right)  ^{2}+\sum_{k>0}\mu_{k}z^{k}%
\end{equation}
These polynomials, for three or more $\mu$s, would seem to be a natural
generalization of those constructed by Neumann, Gegenbauer, and Erd\'{e}lyi.
\ Acting with $H-n^{2}$ gives%
\begin{equation}
\left(  z\frac{d}{dz}\right)  ^{2}\chi_{n}\left(  z\right)  -n^{2}\chi
_{n}\left(  z\right)  +\sum_{k>0}\mu_{k}z^{k}\chi_{n}\left(  z\right)
=\frac{1}{z^{n}}\sum_{j=1}^{n}\left(  \left(  j-n\right)  ^{2}-n^{2}\right)
c_{n,j}z^{j}+\sum_{k>0}\mu_{k}\frac{z^{k}}{z^{n}}\sum_{j=0}^{n}c_{n,j}z^{j}
\label{HonDuals}%
\end{equation}
We now impose the condition that the RHS involve only positive powers of $z$
so as to obtain an inhomogeneity that will be orthogonal to the span of
$\left\{  z^{n}\ |\ n\geq0\right\}  $ under contour integration $%
%TCIMACRO{\toint }%
%BeginExpansion
{\textstyle\oint}
%EndExpansion
\frac{dz}{z}\cdots$, as in previous examples. \ This leads to $n$ equations
that fix the coefficients $c_{n,k}$ for $k=1,\cdots,n$ in terms of $c_{n,0}$,
the latter being an overall choice of normalization. \ Explicitly, the powers
to be eliminated, and the resulting \textquotedblleft
triangular\textquotedblright\ set of $n$ equations, are%
\begin{equation}%
\begin{array}
[c]{cl}%
z^{-n+1}: & c_{n,1}\left(  \left(  1-n\right)  ^{2}-n^{2}\right)  +\mu
_{1}c_{n,0}=0\\
z^{-n+2}: & c_{n,2}\left(  \left(  2-n\right)  ^{2}-n^{2}\right)  +\mu
_{2}c_{n,0}+\mu_{1}c_{n,1}=0\\
\vdots & \vdots\\
z^{-n+n}: & c_{n,n}\left(  \left(  n-n\right)  ^{2}-n^{2}\right)  +\mu
_{n}c_{n,0}+\mu_{n-1}c_{n,1}+\cdots+\mu_{1}c_{n,n-1}=0
\end{array}
\label{nConditions}%
\end{equation}
These equations can always be solved, sequentially, for $c_{n,k}$ in terms of
$c_{n,0}$ and the $\mu$s, as expressed by the recursion%
\begin{equation}
c_{n,k}=\frac{1}{\left(  2n-k\right)  k}\sum_{j=0}^{k-1}\mu_{k-j}%
c_{n,j}\ ,\ \ \ \ \ k=1,\cdots,n. \label{Recursion}%
\end{equation}
So for a given $n$ the $c_{n,k}$ will depend on $\mu_{j\leq n}$ but not on
$\mu_{j>n}$. \ Indeed, the $c_{n,k}$ are finite polynomials in the $\mu$s, and
therefore non-singular for all $\mu$s.

Thus, up to their individual normalizations, the dual functions are completely
determined along with the inhomogeneous equations that they obey,
without\ (yet!) having to find the eigenfunctions$\ $of $H$. \ The resulting
equation giving the action of $H$ on the dual functions is of the form%
\begin{equation}
\left(  H-n^{2}\right)  \chi_{n}\left(  z\right)  =\sum_{k>0}\gamma_{n,k}z^{k}
\label{DualEqn}%
\end{equation}
where the coefficients $\gamma_{n,k}$ may also be expressed explicitly as
finite polynomials in the $\mu$s,%
\begin{equation}
\gamma_{n,k}=\sum_{j=0}^{n}\mu_{k+n-j}c_{n,j}%
\end{equation}
This result follows from (\ref{HonDuals}) and (\ref{Recursion}). \ Note that
$k$ is \emph{not} necessarily limited by $n$ in $\gamma_{n,k}$. \ The
$\gamma_{n,k}$ will depend on all the $\mu$s, in general, with $k$ taking on
all values up to and including the highest power of $z$ appearing in $H$.

With $\left\{  \chi_{n}\left(  z\right)  \right\}  $ in hand, we may then
proceed to construct $\left\{  \psi_{n}\left(  z\right)  \right\}  $ as
analytic functions around the origin by demanding biorthonormality with
$\left\{  \chi_{n}\left(  z\right)  \right\}  $. \ We take the $\psi_{n}$
functions to be infinite series of positive powers%
\begin{equation}
\psi_{n}\left(  z\right)  =z^{n}\sum_{j=0}^{\infty}a_{n,j}z^{j}
\label{AnalyticPsi}%
\end{equation}
If we assume the requisite convergence of these series, their forms
automatically give $%
%TCIMACRO{\toint }%
%BeginExpansion
{\textstyle\oint}
%EndExpansion
\frac{dz}{z}\chi_{k}\left(  z\right)  \psi_{n}\left(  z\right)  =0$ for $n>k$.
\ Otherwise, conditions are obtained on the coefficients $a_{n,j}$ for
$j+n\leq k$ by requiring $\frac{1}{2\pi i}%
%TCIMACRO{\toint }%
%BeginExpansion
{\textstyle\oint}
%EndExpansion
\frac{dz}{z}\chi_{k}\left(  z\right)  \psi_{n}\left(  z\right)  =\delta_{kn}$
for every $k\geq n$. \ These conditions amount to another triangular set of
equations which can always be solved, sequentially, for the $a_{n,j}$, $j\leq
k-n$, for any choice of $\mu$s. \ Namely%
\begin{equation}
a_{n,0}=\frac{1}{c_{n,0}}\ ,\ \ \ \sum_{j=0}^{k-n}c_{k,k-n-j}a_{n,j}=0\text{
\ \ for \ \ }k>n \label{aTriangular}%
\end{equation}
By considering all $k>n$ in succession, we obtain all $a_{n,j}$. \ The series
for $\psi_{n}$\ is a development in the minors that invert these triangular
equations. \ For convenience we choose $c_{n,0}=1$ for all $n$, then%
\begin{align}
\chi_{n}\left(  z\right)   &  =\frac{1}{z^{n}}\left(  1+c_{n,1}z+c_{n,2}%
z^{2}+\cdots+c_{n,n}z^{n}\right) \\
\psi_{n}\left(  z\right)   &  =z^{n}-c_{n+1,1}z^{n+1}+\left\vert
\begin{array}
[c]{cc}%
c_{n+1,1} & 1\\
c_{n+2,2} & c_{n+2,1}%
\end{array}
\right\vert z^{n+2}-\left\vert
\begin{array}
[c]{ccc}%
c_{n+1,1} & 1 & 0\\
c_{n+2,2} & c_{n+2,1} & 1\\
c_{n+3,3} & c_{n+3,2} & c_{n+3,1}%
\end{array}
\right\vert z^{n+3}+\cdots\label{PsiSeries}%
\end{align}
The coefficients of $z^{n+k}$\ in $\psi_{n}\left(  z\right)  $\ are again
finite polynomials in the $\mu$s. \ While convergence of this series is
certainly not obvious for arbitrary $\mu$s, as written, it is clear that
convergence can be determined on a case-by-case basis from the explicit form
of the coefficients. \ Moreover, when the \emph{number} of $\mu$s is finite,
no matter what their values are, it is not too difficult to show the $\psi
_{n}\left(  z\right)  $ are entire functions of $z$.

Thus, if we assume the requisite convergence of the $\psi_{n}$ series, all the
$\left\{  \psi_{n}\left(  z\right)  \ |\ \text{for }n\geq0\right\}  $ are
determined, and they are complete on the span of $\left\{  z^{n}%
\ |\ n\geq0\right\}  $. \ All positive powers of $z$ can be expressed as
series of $\left\{  \psi_{n}\left(  z\right)  \right\}  $, similar in form to
(\ref{Schlomilch}), just as all negative powers of $z$ can be expressed as
finite sums of $\left\{  \chi_{n}\left(  z\right)  \right\}  $, similar in
form to (\ref{PowersExpandedInAs}).

Remarkably, the $\left\{  \psi_{n}\left(  z\right)  \right\}  $ just obtained
turn out to be non-degenerate energy eigenfunctions, and are in fact
\emph{all} of the $2\pi$-periodic eigenfunctions of $H$, with \
\begin{equation}
H\psi_{n}\left(  z\right)  =n^{2}\psi_{n}\left(  z\right)  \label{HEigens}%
\end{equation}
The RHS of (\ref{HEigens}) can be deduced in a novel way from (\ref{DualEqn}),
the biorthogonality of $\left\{  \chi_{j}\left(  z\right)  ,\psi_{k}\left(
z\right)  \right\}  $, and the completeness of $\left\{  \psi_{n}\left(
z\right)  \right\}  $ for analytic functions about the origin, as follows.
\ Given that $\psi_{n}\left(  z\right)  $ is such an analytic function, as in
(\ref{AnalyticPsi}), completeness allows us to write $H\psi_{n}\left(
z\right)  =\sum_{k=n}^{\infty}b_{n,k}\psi_{k}\left(  z\right)  $. \ From this
and biorthonormality, we have $b_{n,k}=\frac{1}{2\pi i}%
%TCIMACRO{\toint }%
%BeginExpansion
{\textstyle\oint}
%EndExpansion
\frac{dz}{z}\chi_{k}\left(  z\right)  H\psi_{n}\left(  z\right)  $. $\ $But
then, upon integrating by parts and using (\ref{DualEqn}) as well as
orthogonality relations, we also have $\frac{1}{2\pi i}%
%TCIMACRO{\toint }%
%BeginExpansion
{\textstyle\oint}
%EndExpansion
\frac{dz}{z}\chi_{k}\left(  z\right)  H\psi_{n}\left(  z\right)  =n^{2}%
\delta_{k,n}$. \ So $b_{n,k}=n^{2}\delta_{k,n}$, and (\ref{HEigens}) is
obtained.\footnote{Note that $z=0$ is a regular singular point of
(\ref{HEigens}), for any number and any values of the $\mu$s such that
$\sum_{k>0}\mu_{k}z^{k}$ converges near the origin. \ In fact,
(\ref{PsiSeries}) is exactly the conventional series obtained by expanding
about the regular singular point at $z=0$, albeit the series was obtained here
in an unusual way from the properties of the dual space polynomials.
\ Convergence criteria for (\ref{PsiSeries}) in terms of the $\mu$s are
obtained in the conventional series approach in \cite{Rofe,Gasymov}, without
any discussion of the dual polynomials.} \ 

Conversely, given (\ref{HEigens}) and (\ref{DualEqn}), we may prove the
orthogonality relations $%
%TCIMACRO{\toint }%
%BeginExpansion
{\textstyle\oint}
%EndExpansion
\frac{dz}{z}\chi_{k}\left(  z\right)  \psi_{n}\left(  z\right)  =0$ for $k\neq
n$ just by inserting $H$ and following the same line of argument as in the
case of a single exponential potential, (\ref{HActingOnDuals}) et seq. \ Some
explicit closed forms for the various functions, some kernels constructed from
sums of their bilinears, and related convergence issues, as well as many other
details for this general case, will be discussed elsewhere.

\section{A brief look at QFT extensions}

For the complex Liouville quantum field theory in 1+1 spacetime, we choose to
define the Hamiltonian density by normal-ordering at mass $m$ in the standard
way \cite{Coleman}. \
\begin{equation}
\mathcal{H}=N_{m}\left(  \frac{1}{2}\pi_{\psi}^{2}+\frac{1}{2}\left(
\partial_{\sigma}\psi\right)  ^{2}+\mu e^{2i\beta\psi}\right)
\label{LiouvilleEnergyDensity}%
\end{equation}
Normal-ordering at any other mass should not affect our conclusions. \ The
resulting (pseudo)scalar field theory is conformally covariant
\cite{CurtrightThorn,BCGTConformal}, at least for a range of $\beta$. \ For
this field theory, the only issue to be raised here is the stability of the
ground state. \ Hollowood \cite{Hollowood} has previously discussed separate
but related issues in his study of solitons for complex Toda theories using
conventional Hilbert space methods and perturbation theory.

We may attempt to place an upper bound on the ground state $\left\langle
\mathcal{H}\right\rangle $ using the conventional Hilbert space norm through
an adaptation \cite{Curtright} of the variational argument of \S III in
Coleman \cite{Coleman}. \ Take as a trial state the zero-mode-shifted Fock
vacuum for arbitrary mass $M$, defined by the effect of annihilation operators
for that mass: \
\begin{equation}
a\left(  0,M\right)  \left\vert 0,\xi,M\right\rangle =\tfrac{1}{2}%
\xi\left\vert 0,\xi,M\right\rangle \ ,\ \ \ a\left(  k,M\right)  \left\vert
0,\xi,M\right\rangle =0\ \ \ \text{if}\ \ \ k\neq0
\end{equation}
For $\left\vert 0,\xi,M\right\rangle $ the zero-mode configuration is
coherent, while the $a\left(  k,m\neq M\right)  $ non-zero mode configuration
is superfluidic (i.e. built formally from powers of correlated pairs of
creation operators $\left(  a^{\dag}\left(  k,m\right)  a^{\dag}\left(
-k,m\right)  \right)  ^{n}$ acting on $\left\vert 0,0,m\right\rangle $).
\ Note that
\begin{equation}
\left\langle 0,\xi,M\right\vert N_{m}\left(  e^{2i\beta\psi}\right)
\left\vert 0,\xi,M\right\rangle =e^{2i\beta\xi}\left\langle 0,\xi,M\right\vert
N_{m}\left(  \cos\left(  2\beta\left(  \psi-\xi\right)  \right)  \right)
\left\vert 0,\xi,M\right\rangle
\end{equation}
since $\left\langle 0,\xi,M\right\vert N_{m}\left(  \sin\left(  2\beta\left(
\psi-\xi\right)  \right)  \right)  \left\vert 0,\xi,M\right\rangle =0$. \ For
this trial state, we find%
\begin{equation}
\left\langle 0,\xi,M\right\vert \mathcal{H}\left\vert 0,\xi,M\right\rangle
=\mu e^{2i\beta\xi}\left(  \frac{M^{2}}{m^{2}}\right)  ^{\beta^{2}/2\pi}%
+\frac{1}{8\pi}\left(  M^{2}-m^{2}\right)  \label{TrialAverageH}%
\end{equation}
Therefore, by choosing $\xi\ $such that $\mu e^{2i\beta\xi}$ is real and
negative, and then letting $M/m\rightarrow\infty$, there is no lower bound on
$\left\langle 0,\xi,M\right\vert \mathcal{H}\left\vert 0,\xi,M\right\rangle $
if $\beta^{2}>2\pi$. \ Thus in the usual Fock space framework, there is no
ground state if $\beta^{2}$ exceeds $2\pi$. \ This may indicate a
phase-transition for the field theory.

For the field theoretic extension corresponding to the model (\ref{H}), with
$\nu=0$, the RHS of (\ref{TrialAverageH}) becomes just a sum over $k=2\beta$
with various $\mu_{k}$. \ Thus the variational argument shows an instability
whenever the sum over $k$\ includes $k>5$.

The problem is now to determine how this variational argument is effected by
using the bilocal norm instead of the usual Hilbert space norm. \ The operator
methods in \cite{BCTExact}\ and/or the functional methods in
\cite{CurtrightGhandour}\ should suffice for this purpose. \ There have been
studies of the variational principle for similar \textquotedblleft
quasi-hermitian\textquotedblright\ theories \cite{Scholtz} and the techniques
introduced in those studies should be applicable to expectations of
(\ref{LiouvilleEnergyDensity}).\ \ This issue will be fully discussed elsewhere.

Here we only remark that the situation is reminiscent of $-\phi^{4}$ field
theory, which also has no ground state based on conventional variational
arguments using the standard Hilbert space norm and trial states with real
expectations $\left\langle \phi\right\rangle $, but which appears to be an
acceptable model when viewed as PT symmetric $\phi^{2}\left(  i\phi\right)
^{2}$ theory \cite{BenderPhi4}. \ This is unquestionably related to the use of
the CPT norm, but more directly, it is due to the definition of the zero-mode
problem on a contour in the LHP, and not the real axis. \ This obviates most
physical intuition about the ground state of that theory, in our opinion.
\ Even so, there are obviously major differences between PT symmetric
$-\phi^{4}$ field theory and imaginary Liouville theory as defined by
(\ref{LiouvilleEnergyDensity}), in that the apparent instability in $-\phi
^{4}$ field theory\ is evident at the classical level for any coupling
strength, whereas the instability indicated by (\ref{TrialAverageH}) for
$\beta^{2}>2\pi$ is a purely quantum effect.

\section{Conclusions}

We made a thorough analysis of some complex one dimensional periodic
potentials, of Liouville type, within the framework of biorthogonal systems.
\ We provided an extensive guide to the considerable literature for these
models. \ We conclude that these systems are consistent at both classical and
quantum mechanical levels, with remarkable and deep connections between the
two levels, just as in the case of real Liouville models. \ In fact, the
connections are more intimate in the complex case, inasmuch as the classical
structure is richer. \ We studied in some detail the correspondence and impact
of the classical dynamics on that of the quantum problem. \ 

More specifically we have shown that the complex Liouville model is equivalent
to a projected free particle system, and we may choose that free particle
projection to be chiral. \ With that choice, the complex Hamiltonian may be
viewed as a completely local description for the free chiral particle, the
only such local Hamiltonian description known to us.

We have used the remarkable properties of Neumann polynomials as the
biorthogonal partners of Bessel functions on a circle in the complex plane.
With those properties in hand, we obtained closed forms for the inner product
and various other hermitian kernels. \ We enlarged this study to include a
wide class of complex periodic one dimensional Hamiltonians, and we obtained
similar closed form results in several additional cases. \ 

In our analysis, the theory of PT symmetric models was often a guide but did
not play a decisive role insofar as the states were self-orthogonal in the
periodic sectors that we emphasized and some of the models that we considered
were not PT symmetric. \ Moreover, many of the concepts that we used are
already contained in the previously existing framework of biorthogonal
systems. \ As stressed in the title of the paper, the use of biorthonormal
systems of vectors for the construction of consistent quantum mechanical
systems was the real conceptual lift, applicable even when the states are
self-orthogonal.\ \ The most important feature shared with the PT symmetric
approach was simply that of a complex Hamiltonian with real eigenvalues.

It would be interesting to investigate biorthogonal systems for
self-orthogonal subsectors of other solvable complex potential models, e.g.
those in \cite{Khare}, for which the duals would seem to be the associated
functions described in \cite{Veliev}. \ Another interesting and possibly
important challenge is to combine the biorthogonal approach with the study of
symmetries of the Hamiltonian. \ Beyond the usual commutators between any
local symmetry generators, there must also be commutation relations among the
various bilocal kernels. \ Thus the full set of symmetries of the
corresponding quantum problem may not be local and may not be obvious. \ Along
these lines, it is perhaps a useful working hypothesis that any consistent
local complex Hamiltonian with real eigenvalues is related by a similarity
transformation to that of a hermitian, albeit nonlocal, Hamiltonian.
\ However, this conceptual device may not always lead to practical
calculational tools as this similarity transformation is usually not known, a
priori, and may be difficult to implement when it is known.

A clear and readable discussion on the use of similarity transformations to
expose the hermitian underpinnings of certain \emph{apparently} non-hermitian
(i.e. \textquotedblleft quasi-hermitian\textquotedblright) theories is given
in \cite{Scholtz} (also see \cite{MostaJMP6,MostaJPA1,MostaJPA3,MostaBatal}).
\ The general theory is nicely illustrated in a recent paper on PT-symmetric
$-z^{4}$ quantum mechanics \cite{JonesMateo}, where after the change of
variable $z=-2i\sqrt{1+ix}$, $x\in\mathbb{R}$, the relevant similarity
transformation is very simple and easily implemented.

While substantial progress has already been made on several PT symmetric field
theories \cite{BenderReview,MostaReview}, it is important to apply the methods
of biorthogonal systems to more general quantum field theories and many-body
problems, particularly those of the complex Liouville and Toda type. \ It
remains a challenge to consider situations involving self-orthogonal states in
these models, to compute correlation functions using those states, and to
investigate their possible relevance for string theories.

The study of the present paper may lead to a deeper appreciation of the
behavior of quantum Liouville field theory for complex coupling constants, as
sketched in Section 8. This would further understanding of subcritical string
theory, especially the correlators for those models which exhibit interesting
structure in the complex coupling plane (cf. \cite{Nakayama} and references
therein, especially \cite{DornOtto,Zamolodchikov}). \ In fact, Liouville
theory with imaginary coupling (or its close cousin under complex continuation
-- so-called \textquotedblleft timelike\textquotedblright\ Liouville theory)
has already been analyzed using conformal field theory methods
\cite{Strominger,Schomerus}. We believe the Hamiltonian analysis initiated
above will help shed additional light on this subject.

Moreover, the field theory extensions of the models described in Section 7 are
relevant to applications of conformal field theory and phase-transitions. \ In
particular, complex Liouville field theory may be useful to understand
freezing transitions \cite{CarpentierLeDoussal,CastilloLeDoussal}. \ More
recently, there is very interesting work \cite{Affleck} involving complex
Liouville field theory in non-Hermitian Luttinger liquids and flux line
pinning in planar superconductors. \ As for other possible physical
applications of the techniques in this paper, it is perhaps interesting to
conjecture that coupled Bose-Einstein systems with dissipative effects might
be governed by relatively simple non-hermitian effective Hamiltonians. \ This
could be, in several respects, a completely quantum analogue of the classical
driven/damped oscillator systems discussed in \cite{Heiss}.

\paragraph{Acknowledgements}

We thank C Bender for introducing us to PT symmetric theories and raising our
awareness of non-hermitian Hamiltonian methods. \ We also thank P G O Freund,
A Veitia, and D Schuster for useful discussions. \ One of us (TC) thanks the
Aspen Center for Physics and the Institute for Advanced Study for providing
stimulating environments in which this work was completed. \ Finally, we thank
the anonymous referee for making helpful suggestions and for bringing some
relevant papers to our attention. \ This material is based upon work supported
by the National Science Foundation under Grant No. 0303550.

\end{document}